\documentclass[openbib]{rsproca_new}
\usepackage[]{graphicx}  
\usepackage{mwe,tikz}
\usepackage[percent]{overpic}
\usepackage{subfig}
 
\usepackage{amsmath}       
\usepackage{amsfonts}
\usepackage{bm}
\usepackage{amssymb}
\usepackage[T1]{fontenc}
\usepackage{booktabs}
\usepackage{array}
\usepackage{lipsum}                     
\usepackage{xargs}                      
\renewcommand\newblock{\,}

\def\real{I\!\!R}
\def\k{\mathbf{k}}

\def\gk{\boldsymbol{\kappa}}
\def\gh{\bar{\boldsymbol{\kappa}}}

\def\D{\mathbf{D}}

\newtheorem{theorem}{\bf Theorem}[section]

\newtheorem{proposition}[theorem]{Proposition}

\graphicspath{{./figures/}}
\pdfoutput=1
\begin{document}

\title{Multiparameter actuation of a neutrally-stable shell: a flexible gear-less motor}
\author{W.~Hamouche$^1$, C.~Maurini$^1$, S.~Vidoli$^2$, A.~Vincenti$^1$}
\address{$^1$Sorbonne Universit\'{e}s, UPMC Univ Paris 06, CNRS, UMR 7190, Institut Jean Le Rond d'Alembert, F-75005 Paris, France \\
$^2$Sapienza Universit\`a di Roma, Dip. di Ingegneria Strutturale e Geotecnica, \\
Via Eudossiana 18, 00184 Rome, Italy}
\subject{Structural engineering, Mechanics, Mathematical modelling }
\corres{Insert corresponding author name\\
\email{corrado.maurini@upmc.fr}}
\begin{abstract}
We have designed and tested experimentally a morphing structure  consisting of a neutrally stable thin cylindrical shell driven by a multiparameter piezoelectric actuation. The shell is obtained by plastically deforming an initially flat copper disk, so as to induce large isotropic and almost uniform inelastic curvatures. Following the plastic deformation, in a perfectly isotropic system, the shell is theoretically neutrally stable, owning a continuous manifold of stable cylindrical shapes corresponding to the rotation of the axis of maximal curvature. Small imperfections render the actual structure bistable, giving preferred orientations. A three-parameter piezoelectric actuation, exerted through micro-fiber-composite actuators, allows us to add a small perturbation to the plastic inelastic curvature and to  control the direction of  maximal curvature.  This actuation law is designed through a geometrical analogy based on  a fully non-linear inextensible uniform-curvature shell model.  We report on the fabrication, identification, and experimental testing of a prototype and demonstrate  the  effectiveness of the piezoelectric actuators in controlling its shape. The resulting motion is an apparent rotation of the shell, controlled by the voltages as in a "gear-less motor", which is, in reality, a precession of the axis of principal curvature. \end{abstract}


\maketitle
\section{Introduction}
Thin structures, such as plates and shells, are essentially surfaces that can stretch and bend. 
Stretching, \emph{i.e.} the change of metric of the midplane, requires that the material is deformed uniformly through-the-thickness and that the elastic energy is proportional to  the shell thickness $h$. Bending, \emph{i.e.} the change of curvature of the midplane, implies a  material deformation linear through-the-thickness and an elastic energy proportional to $h^3$.
Moreover, the metric and the curvature of a surface must obey a fundamental compatibility equation that relates the change of the Gaussian curvature (i.e. the product of the principal curvatures) to the in-plane stretching \cite{Cia05}. In the differential geometry of surfaces this compatibility equation stems from the Gauss Theorem. 
 The combination of these effects gives thin shells unique features:
(i)  soft inextensional modes  at constant  Gaussian curvature;
(ii) stiff modes implying a stretching of the
mid-surface, i.e. a change of the Gaussian curvature;
(iii) possible transition paths between several stable equilibria (minimal energy configurations) characterised by large shape-changes, almost constant Gaussian curvature, and small material deformations. 
 Researchers and engineers exploit these properties to design shape-changing reconfigurable (or morphing) structures where thermal effects \cite{SefMcM07}, swelling \cite{EfrShaKup09a}, piezoelectricity \cite{BowButJer07,SchHye03,PorCamWea08,FerMauVid10a,LeeMooInm17} or photoelasticity  \cite{NaBenBae16} can be used to control the shape. All these effects introduce inelastic deformations, prescribing target values for the metric and curvature of the surfaces  \cite{ModBhaWar11,SefMau13,NaBenBae16,KleEfrSha07,PezSmiNar16}. If these target deformations are not compatible, equilibrium configurations are obtained through an elastic misfit and are pre-stressed. 
Lewicka and coworkers \cite{LewMahPak14} recently developed a rigorous asymptotic analysis of non-Euclidean plates \cite{EfrShaKup09a}  with geometrically incompatible inelastic deformations, assessing the appropriate two-dimensional models and the possible energy scaling regimes \cite{LewMahPak14} using dimensional reduction techniques based on the direct methods of the calculus of variations \cite{FriJamMul06}.\\

In this paper, we  exploit the effect of incompatible inelastic deformations to design shells where a weak embedded actuation drives large shape-changes.  Previously,  several authors \cite{BowButJer07,SchHye03,PorCamWea08,LeeMooInm17} tried to use piezoelectric actuators to trigger snap-through instabilities between two stable equilibrium shapes of bistable composites cross-ply laminated plates, mimicking the behaviour of a Venus Flytrap's leaf \cite{ForSkoDum05}. Here, we draw on a different concept: we  design a structure characterised by several stable equilibria separated by vanishing energy gaps, similar to the neutrally stable (or  zero-stiffness) shells of \cite{guest2014}. Hence we are able to obtain large shape-changes with low actuation efforts and to continuously drive the transition between different stable equilibria.
As proposed in \cite{GalGue04,guest2014}, such a neutrally stable shell is built by introducing isotropic  plastic curvatures in isotropic initially flat disk. This generates  cylindrical shells with a special zero-stiffness "mode", where the axis of curvature can rotate freely. Our idea is to use a set of surface-bonded piezoelectric actuators to apply a perturbation to the initial plastic curvature, so as to continuously control the directions of principal curvature of the shell. Here we elaborate on the modelling and design principles. We also report on the fabrication and test of an experimental prototype, proving the practical effectiveness of the proposed concepts. \\

To introduce our goals, in Section~\ref{sec:overview} we outline our final experimental result. 
In Section~\ref{sect:0stiff}, we present a simple nonlinear model obtained assuming that the shell is inextensible and the curvature fields uniform. We show that, in such a framework, the search of the stable equilibria of the shell for a given inelastic curvature is equivalent to finding the points on a cone with minimal distance from a given point of $\real^3$.
This equivalence is sufficient to derive, in Section~\ref{sect:act}, the multiparameter actuation strategy to produce a complete $360^{\circ}$ precession of the shell curvature.
In Section~\ref{sect:exp} we report in detail the experimental findings, comparing them with the predictions of the theoretical model and finite element results.

\section{Overview of the main experimental result}
\label{sec:overview}

\begin{figure}[ht]
\centering
\begin{tabular}{ccc} 
\includegraphics[width=.475\textwidth]{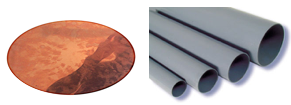}&
\includegraphics[width=.225\textwidth]{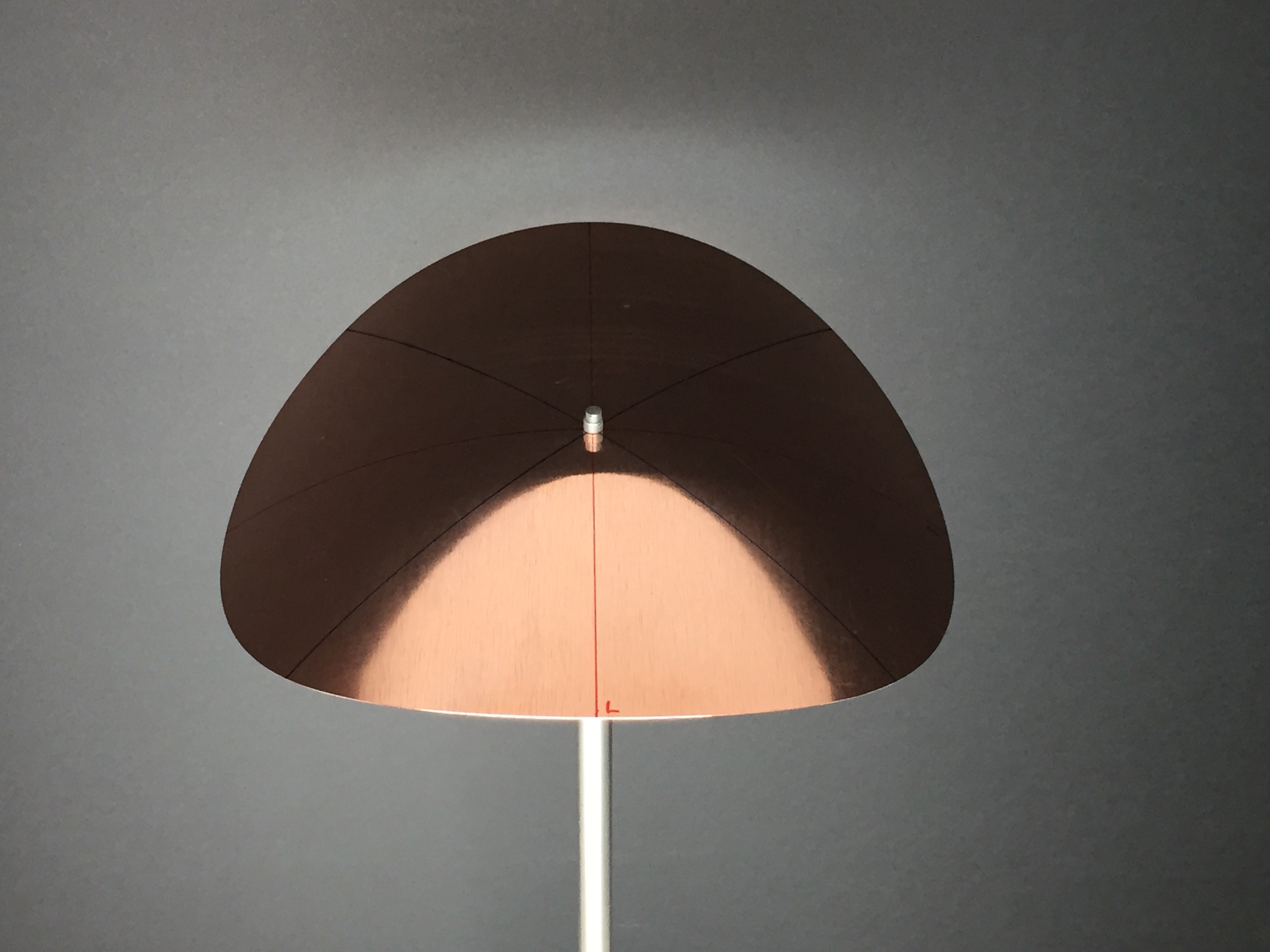} &
\includegraphics[width=.225\textwidth]{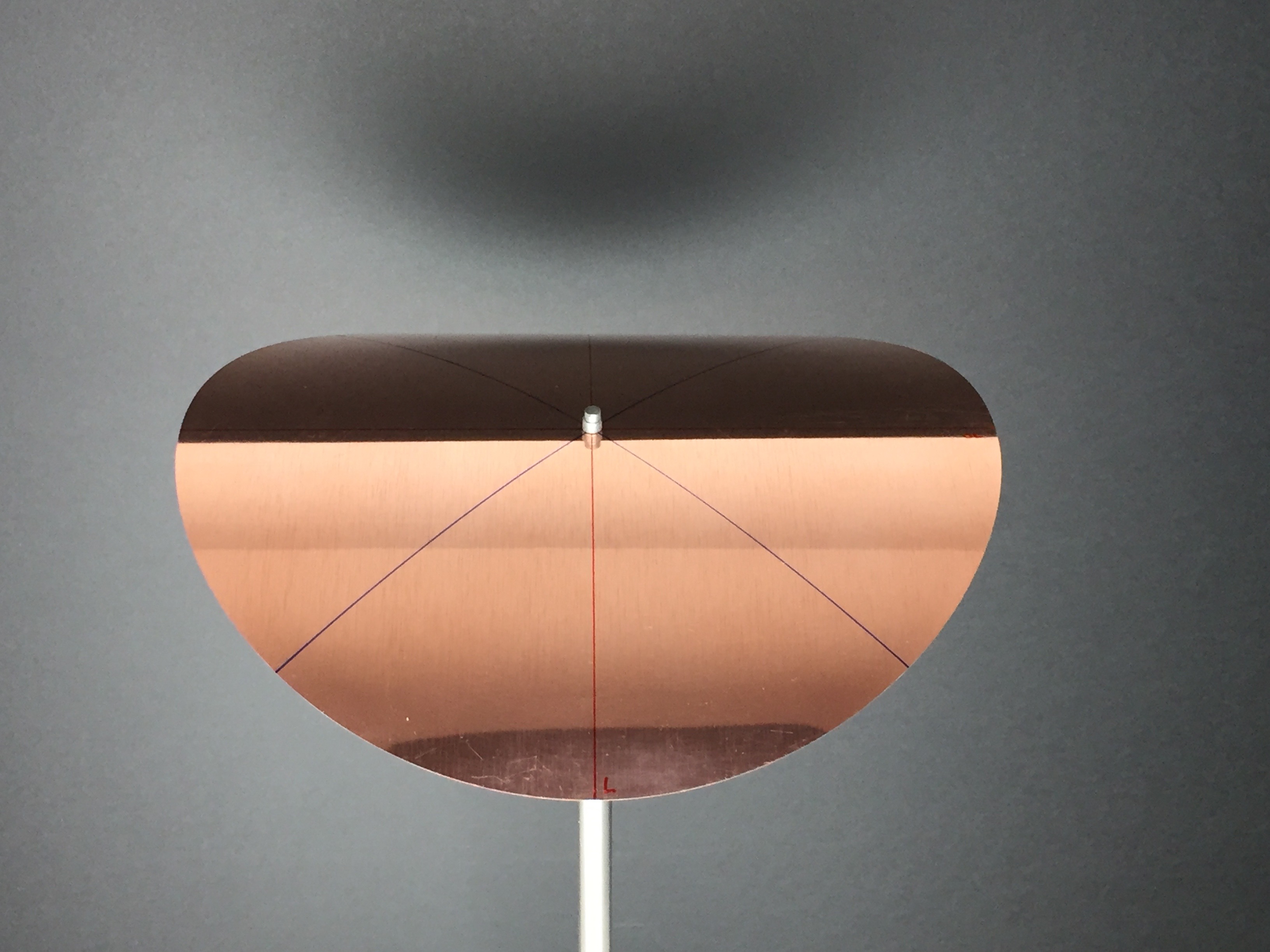}  \\
(a) & (b) & (c) 
\end{tabular}
\caption{An initially flat stress-free disk of radius $L=122.5\,\text{mm}$ and thickness $h=0.3\,\text{mm}$  is winded over  PVC cylinders  (a)  of progressively smaller diameters ($160\,\text{mm}$, $140\,\text{mm}$, and   $110\,\text{mm}$) to generate equal plastic curvatures in two orthogonal directions (isotropic inelastic curvature). The resulting structure is bistable. The first (b) and second (c) stable configurations are approximately cylindrical, with principal curvature axes rotated of $\pi/2$.}
\label{fig:pla}
\end{figure}

We fabricate a cylindrical  shell by  manually winding an initially flat thin copper disk around cylinders of progressively decreasing diameters, see Figure~\ref{fig:pla}. The procedure is repeated several times, by plastically bending the shell along two mutually orthogonal directions. We apply  inelastic curvatures with a radius of curvature of the order of the diameter of the shell. In this regime,  geometrically non-linearities play an important role, and the behaviour of the shell is approximately inextensible. Inextensibility implies that the equilibrium configurations are cylinders. The plastic deformation process is modelled by  an  isotropic and homogenous tensor of inelastic curvature  in the form $\bar \k= \bar {k}_{\textsc{P}} \,\mathbf{I}$, where $\mathbf{I}$ is the two-by-two identity matrix, and $\bar {k}_{\textsc{P}}$ gives the amplitude of the inelastic curvature of plastic origin.  In the perfectly isotropic case, theoretical models predict that the shell is \emph{neutrally stable}:  all the cylindrical configurations having an approximately uniform and uniaxial curvature tensor in the form $\k=k \,\bold{e}(\varphi)\otimes\bold{e}(\varphi)$ have then same elastic energy,  $\bold{e}(\varphi)$ being a unit vector oriented with an angle $\varphi$ (\emph{e.g.} see Figure~\ref{fig:conecoords}b). These configurations constitute a \emph{zero-stiffness} mode of neutrally-stable equilibria, as shown in \cite{GalGue04,guest2014}. In practice, imperfections, which we attribute mainly to anisotropic plastic hardening, render the disk \emph{bistable} at the end of the plastification phase. Indeed, as pointed out in \cite{SefGue11}, neutral stability is a singular regime, which is not robust with respect to small perturbations of the system parameters. 
 However, since the energy required to rotate the axis of principal curvature is very low, the same shell can be classified as \emph{almost} neutrally stable. 

\begin{figure}[ht]
\centering
\includegraphics[width=.5\textwidth]{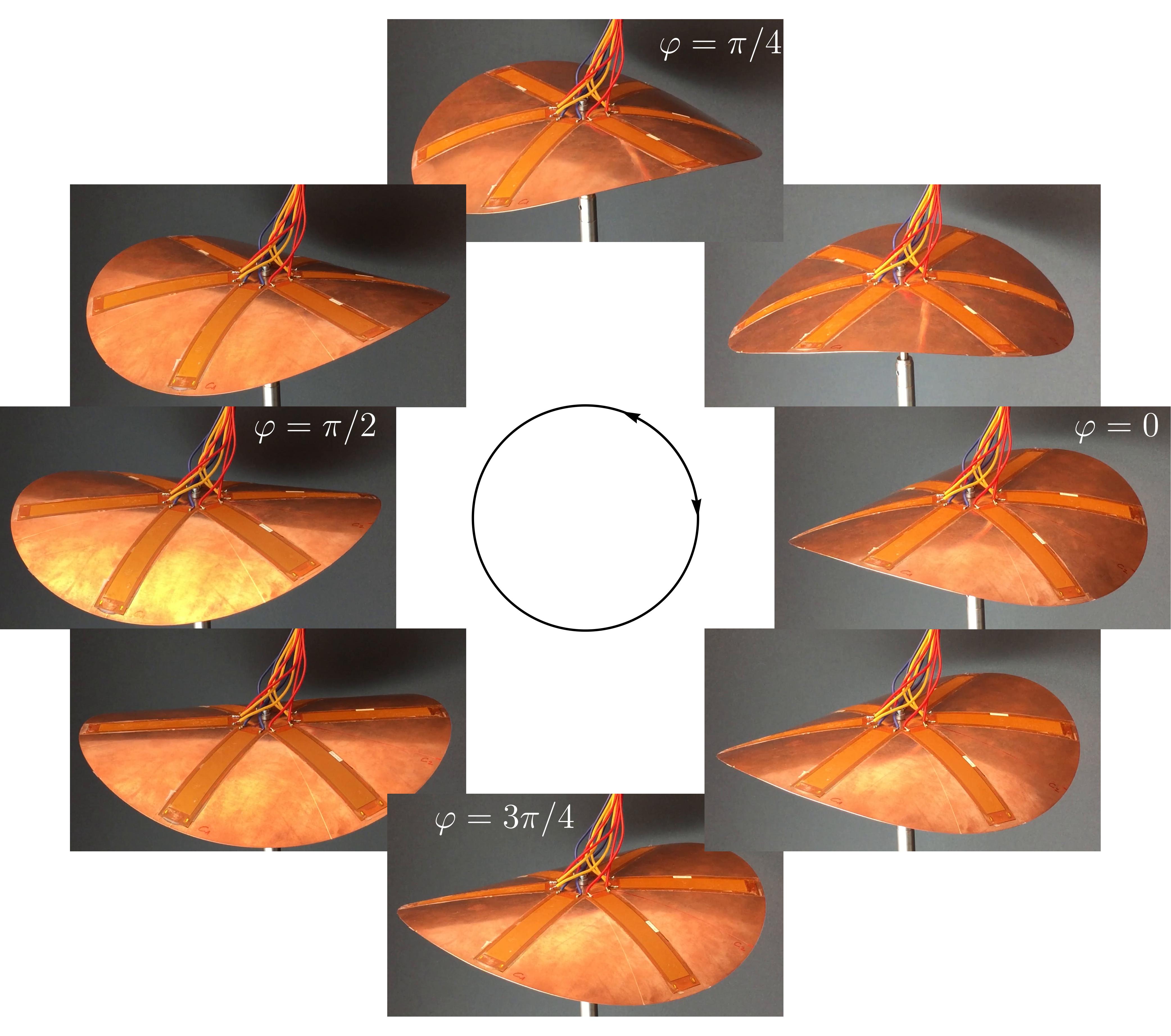}
\caption{Sequence of shell configurations during the actuation showing a complete precession of the axis of maximal curvature. The images are frames of a Video provided in  the Supplementary Material.}
\label{fig:sequence}
\end{figure}

After plastification, on the top surface of the shell, we bond three pairs of piezoelectric Macro-Fiber-Composite actuators (MFC) aligned along three directions, which are mutually rotated by an angle $120^\circ$. Each MFC pair  is connected in parallel and driven by an independent actuation voltage, whose effect is to introduce a small inelastic curvature of piezoelectric origin in the corresponding direction. Driving the three pairs of MFC  with an appropriate phase-shift introduces an equivalent uniaxial piezoelectrically-induced inelastic curvature $\bar{\mathbf{k}}_{\textsc{V}}$, which is very small with respect to that of plastic origin. We show that by varying the voltages so as to rotate the orientation of $\bar{\mathbf{k}}_{\textsc{V}}$, the principal axis of the cylindrical equilibrium shape rotates accordingly (see Figure~\ref{fig:sequence}):  we have obtained a  \emph{flexible gear-less motor}. The most effective illustration of this result is given by the video provided as a Supplementary Material. The apparent $360^\circ$rotation of the shell is actually a deformation process at almost constant elastic energy characterised by a precession of the axis of maximal curvature.

\section{Inextensible shells with uniform curvature}\label{sect:0stiff}

We start introducing a simplified model for the shell, which  builds on those presented in \cite{Sef07,VidMau08,GalGue04}. It is based on the following basic assumptions, which are satisfied in our case: (i) the shell is inextensible and initially flat; (ii) the shell curvature is uniform in space; (iii) the material is orthotropic and linearly elastic, (iv) the shell is free on the boundary, and the loading is given only by imposed inelastic curvatures. 
The inelastic curvatures are used to model the plastic deformations and the effect of the embedded piezoelectric actuators: they are supposed to be given data. We consider an orthotropic material behaviour because, as we will show later, the disk after plastification is not perfectly isotropic. 
The model is  geometrically fully nonlinear, \emph{i.e.} the curvature, despite being supposed uniform in space, can be arbitrarily large. 
We show that with these hypotheses, the problem of finding the stable equilibria has a suggestive and useful geometrical analogy. Namely, we show that it is equivalent to finding the points on a cone (the inextensibility constraint for the curvatures) having a minimal distance from a given target point of $\real^3$ (the components of the inelastic curvature tensor). Moreover, we will  reduce this problem to finding the minima of a function of a single scalar variable.  
 
\subsection{Governing equations for inextensible Koiter shells and stable equilibria}\label{sect:model}

We take as reference configuration  for the shell the flat disk $\Omega=\{(x_1,x_2)\in\real^2, x_1^2+x_2^2\leq L^2\}$, where we  $(x_1,x_2)$ are cartesian coordinates and $L$ being the radius of the disk.
We denote by $x\in \Omega$ and $y=f(x)\in \mathcal{S}\subset\real^3$, respectively, the generic point in the reference configuration and its placement after the transformation,  $\mathcal{S}=f(\Omega)$ being the curved current configuration of the shell.
According to standard differential geometry, denoting by $\partial_\alpha f=\partial f /\partial x_\alpha$, and using repeated index notation with $\alpha,\beta=1,2$, the metric and the curvature of the deformed surface $\mathcal S$ are given by 
\begin{equation}
a_{\alpha\beta}(f)=\partial_\alpha f\cdot \partial_\beta f,\qquad
b_{\alpha\beta}(f)= \partial_{\alpha\beta} f\cdot \frac{\partial_\alpha f\times \partial_\beta f}{\vert\partial_\alpha f\times \partial_\beta f\vert},
\end{equation}
 where $\cdot$ and $\times $ denotes the scalar and vector products. The interested reader is referred to \cite{Cia05} for more details. 
 
The first and second fundamental forms in the flat reference configuration are 
\begin{equation}
a_{0\alpha\beta}=\delta_{\alpha\beta}, \quad b_{0\alpha\beta}=0, 
\qquad \alpha,\beta=1,2,
\end{equation}
where $\delta_{\alpha\beta}$ means the Kronecker delta.
The relevant measures of deformation of a Koiter shell model are  the change-in-metric tensor $e$ (membrane deformation) and the change-in-curvature tensor $k$ (bending deformation). Their covariant components read respectively:
\begin{equation}
e_{\alpha\beta}(f)=\dfrac{1}{2}\,(a_{\alpha\beta}(f)-a_{0\alpha\beta})\equiv \dfrac{1}{2}\,(a_{\alpha\beta}(f)-\delta_{\alpha\beta}),\qquad
k_{\alpha\beta}(f)=b_{\alpha\beta}(f)-b_{0\alpha\beta}\equiv b_{\alpha\beta}(f).
\end{equation}

As anticipated, we consider an inextensible shell model, assuming a vanishing variation of the metric tensor in every  point, namely $e_{\alpha\beta}(f)=0$ in $\Omega$. Hence, assuming a linearly elastic material behaviour, the only contribution to the elastic energy  is the following quadratic form of the curvature tensor
\begin{equation}
\mathcal{E}_b=\dfrac{1}{2} \int_{\Omega} D^{\alpha\beta\sigma\tau} \,(k_{\sigma\tau}(f) - \bar k_{\sigma\tau})\,(k_{\alpha\beta}(f)- \bar k_{\alpha\beta}) \,d\Omega,
\label{bendingen}
\end{equation}
where $D^{\alpha\beta\sigma\tau}$ are the contravariant components of the bending stiffness tensor and $\bar k_{\alpha\beta}$ are the covariant components of a symmetric tensor representing the inelastic, or \emph{target}, curvatures. The inelastic curvature, or their associated inelastic pre-stresses $m^{\sigma\tau}=-D^{\alpha\beta\sigma\tau}\,\bar k_{\alpha\beta}$, may model different physical effects: thermal and hygroscopic effects, plastic deformations or actuation by embedded active materials, as piezoelectric actuators.

In absence of external loads, the functional \eqref{bendingen}  coincides with the  total potential energy of the shell. In a variational setting, the stable equilibrium configurations  are the solution of following constrained minimisation problem
\begin{equation}
\min \mathcal{E}_b, \text{ among all $f$ such that }  e_{\alpha\beta}(f)=0.
\label{minu}
\end{equation}
 
In the spirit of intrinsic approaches to elasticity \cite{CiaGraMar09}, one could tackle the same problem by minimising the energy in terms of curvature fields which are compatible with inextensible deformations. 
In particular, if $\Omega$ is simply connected, the necessary and sufficient conditions for the field $k_{\alpha\beta}$ to be compatible with inextensible deformations of an initially flat surface with the metric
$a_{\alpha\beta}=a_{0\alpha\beta}=\delta_{\alpha\beta}$  are
\begin{equation}
\mathrm{det}k_{\alpha\beta}=0 \quad\text{and}\quad \partial_{\beta} k_{\alpha\sigma}- \partial_{\sigma} k_{\alpha\beta} =0, \text{ in }\Omega.
\label{comprho}
\end{equation}
These equations are the particularisation of the general Gauss and Codazzi-Mainardi compatibility equations for the surface $\mathcal{S}$, for the case of initially flat shells: they assure necessary and sufficient conditions for the existence of a transformation  $f$ having given metric and curvature fields \cite{Cia05}.

Hence, an intrinsic approach, equivalent to \eqref{minu}, to find the stable equilibria  is  to find the curvature field $k_{\alpha\beta}$ solution of
\begin{equation}
\min \mathcal{E}_b,\quad \text{among all }k_{\alpha\beta} \text{ such that } 
 k_{\alpha\beta}=k_{\beta\alpha},  
\;\; \mathrm{det}k_{\alpha\beta}=0,\;\;  
\ \partial_{\beta} k_{\alpha\sigma}= \partial_{\sigma} k_{\alpha\beta} \text{ in }\Omega.
\label{minrho}
\end{equation}
The fundamental theorem of surface differential geometry \cite{Cia05} assures the equivalence between the solutions of \eqref{minrho} and \eqref{minu}, up to rigid body displacements.

\subsection{The uniform curvature assumption and a geometric analog problem}\label{sect:cone}
We suppose that all the covariant components of both the curvature tensor $k_{\alpha\beta}$ and the inelastic-curvature tensor $\bar{k}_{\alpha\beta}$ are uniform in space. Hence, the Codazzi-Mainardi conditions in \eqref{minrho} are trivially verified. 
After re-organizing these components in the Voigt vectors $\k=(k_{11},k_{22},2k_{12})^\top$ and $\bar{\k}=(\bar{k}_{11},\bar{k}_{22},2 \bar{k}_{12})^\top$, the problem \eqref{minrho}
is reformulated under the assumption of uniform curvatures as
\begin{equation}
\min_{\k\in\real^3} \left[\dfrac{1}{2} \,\D\, (\k-\bar{\k}) \cdot (\k-\bar{\k})\right], \quad\text{ such that }\quad  \mathrm{det}\,\k:=k_{11}k_{22}-k_{12}^2=0,
\label{minrhouc}
\end{equation}
where  $\D$ is a symmetric $3\times 3$ matrix representing the bending stiffness. To deduce its components, from the contravariant components $D^{\alpha\beta\sigma\tau}$ in \eqref{bendingen}, one must perform an average over $\Omega$ and use the Voigt representation of indices.
For orthotropic materials having the coordinate directions $x_1$ and $x_2$ as symmetry planes, the stiffness matrix $\D$ can always be reduced to the form
\begin{equation}
\D:= D\, \left(
\begin{array}{ccc}
1 & \nu & 0 \\
\nu & \beta & 0 \\
0 & 0 & \alpha\,\dfrac{1-\nu}{2}
\end{array}
\right),
\quad D>0,\;\nu^2 <\beta, \; 0< \beta\leq1, \; \alpha>0.
\label{Dmatrix}
\end{equation}
For homogenous shells, $D={S E_{1}h^3}/(12 (1-\nu^2))$, where $S$ is the area of $\Omega$, $E_{1}>0$ is the Young modulus in the coordinate direction $x_1$, $h$ the thickness of the shell, $\nu$ the Poisson ratio, $\beta=E_2/E_1$ the ratio of Young moduli (being $E_2 < E_1$), and $\gamma=\alpha({1-\nu})/{2}$ the shear stiffness.
The conditions in \eqref{Dmatrix} on the  material parameters imply a positive-definite stiffness matrix; for an isotropic material $\beta=\alpha=1$.

\begin{figure}[h!]
\begin{center}
{
\subfloat[]{\includegraphics[width=.4\textwidth]{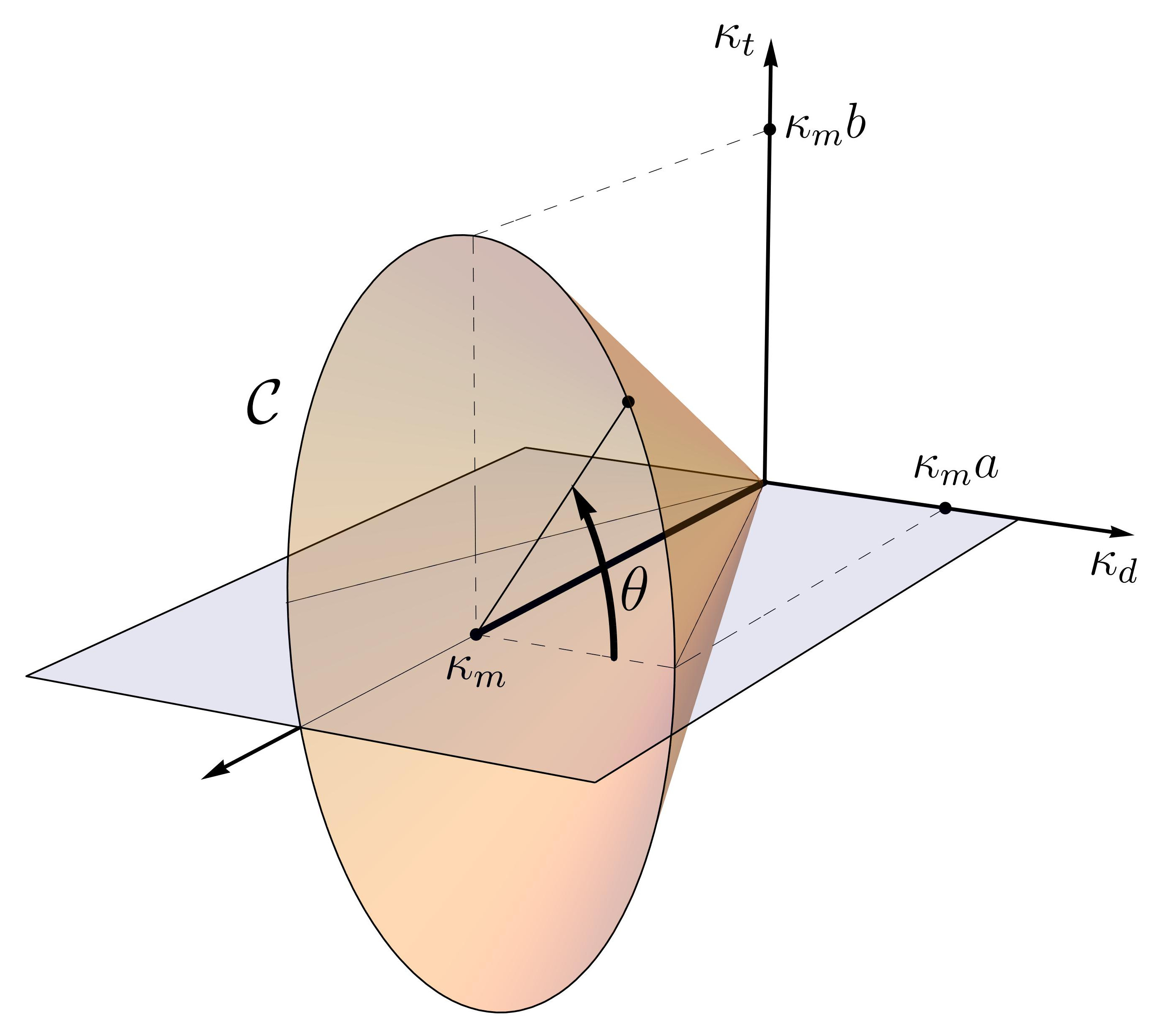}}
\quad
\subfloat[]{\includegraphics[width=.3\textwidth]{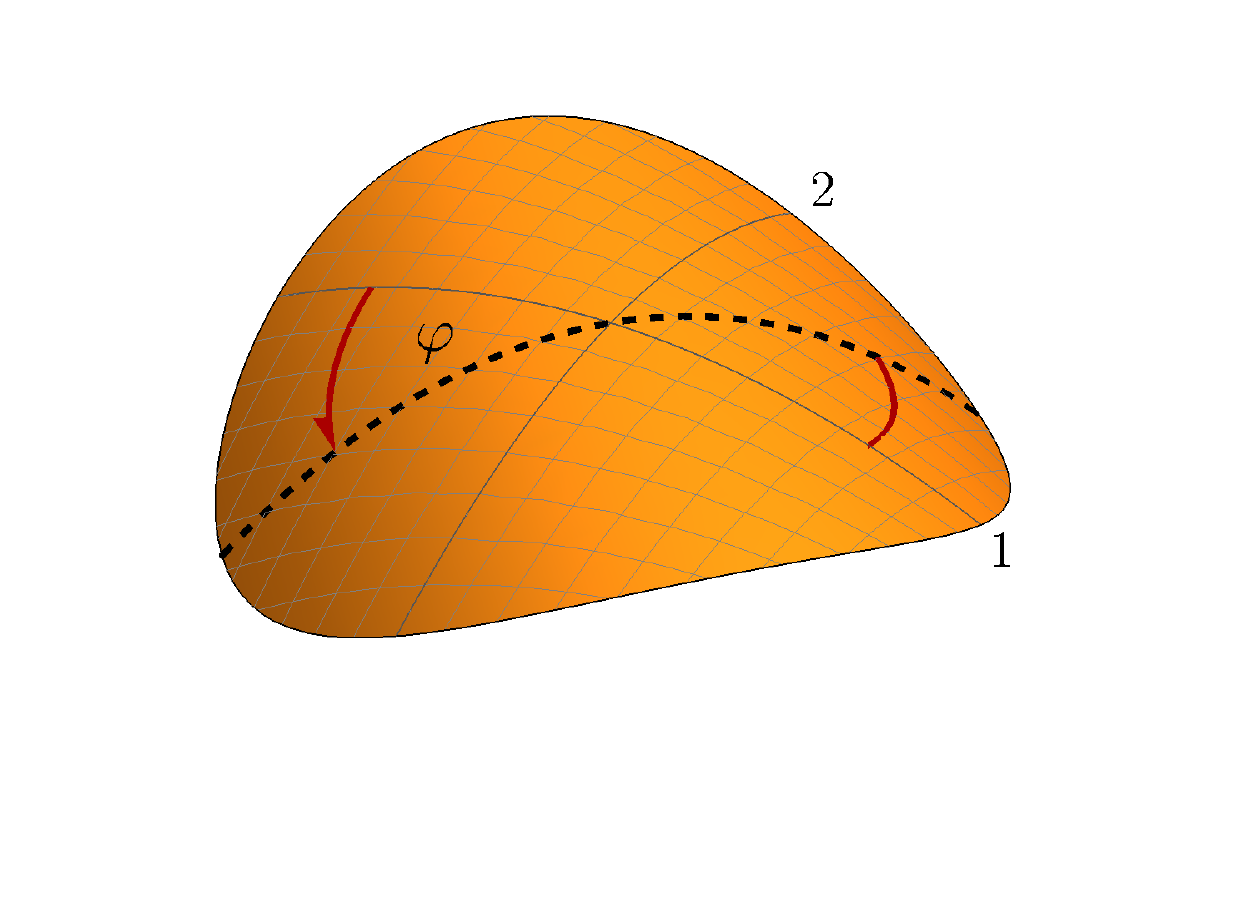} }
}
\caption{(a): Coordinates $(\kappa_m,\theta)$ in \eqref{konC}  on the cone of inextensible curvatures. (b): Corresponding cylindrical shell shape with curvature $\mathbf{k} =k\, \mathbf{e}(\varphi)\otimes \mathbf{e}(\varphi)$ where $k$ and $\varphi$ are given in \eqref{phitheta}. For $\beta=1$, $k=\kappa_m/\sqrt{1+\nu}$ and $\varphi=(\pi-\theta)/2$. }
\label{fig:conecoords}
\end{center}
\end{figure}

The present formulation depends on the three dimensionless material parameters  $\nu$, $\beta$ and $\alpha$. In order to further simplify the formulation and reduce the number of relevant parameters, we introduce the change of coordinates for the curvature tensor $\k$: 
\begin{equation}
\kappa_m=
   \sqrt{1+\dfrac{\nu}{\sqrt{\beta }}}
   \,\left(\dfrac{k_{11}+k_{22}\sqrt{\beta}}{2}\right),\quad
\kappa_d=
   \sqrt{1-\dfrac{\nu }{\sqrt{\beta }}}
   \,\left(\dfrac{k_{22}\sqrt{\beta}-k_{11}}{2}\right),\quad
\kappa_t=\sqrt{(1-\nu)\alpha}\,k_{12}.\quad
\label{ccoorho}
\end{equation}
Expressions \eqref{ccoorho} can also apply to the tensor of inelastic curvatures $\bar \k$ and we define $\gk:=\{\kappa_m,\kappa_d,\kappa_t\}$ and $\bar{\gk}=\{\bar{\kappa}_m,\bar{\kappa}_d,\bar{\kappa}_t\}$. 
As far as the matrix $\D$ is strictly positive definite, these relations are invertible and, therefore, represent a genuine change of coordinates in the space of curvatures. 
The reason for choosing \eqref{ccoorho} is that the quadratic form of the bending energy is diagonalised in terms of coordinates $\gk$ and $\bar{\gk}$ as
\begin{equation}
\mathcal{E}_b = \dfrac{1}{2}\, \D (\k-\bar\k)\cdot (\k-\bar\k)={D}\,\left[ (\kappa_m-\bar\kappa_m)^2+(\kappa_d-\bar\kappa_d)^2+(\kappa_t-\bar\kappa_t)^2\right]= D\,\,\Vert \gk-\bar\gk\Vert^2.
\label{bendiag}
\end{equation}
Moreover, using \eqref{ccoorho}, the condition of inextensibily reads
\begin{equation}
\mathrm{det}\,\k= \dfrac{1}{\sqrt{\beta}+\nu} \,\left( \kappa_m^2- \dfrac{\kappa_d^2}{a^2}-\dfrac{\kappa_t^2}{b^2}\right)=0.
\label{defcone}
\end{equation}
where we have introduced the following two positive non-dimensional parameters: 
\begin{equation}
a=\sqrt{\dfrac{\sqrt{\beta }-\nu}{\sqrt{\beta }+\nu}}, \qquad
b=\sqrt{\alpha\dfrac{1-\nu }{\sqrt{\beta }+\nu }}.
\label{defsemiaxes}
\end{equation}
Hence, in the new system of coordinates, the set of curvatures satisfying the inextensibility constraint \eqref{defcone} is the  cone 
\begin{equation}
\mathcal{C}:=\{\gk\in\real^3: \; c(\gk):={\kappa_d^2}/{a^2}+{\kappa_t^2}/{b^2}- \kappa_m^2=0\}.
\label{conedef}
\end{equation}
The axis of the cone is aligned with the $\kappa_m$-direction. The semi-axes of its  elliptical cross-section ($\kappa_m=\text{const}$) are aligned with  the $\kappa_d$ and $\kappa_t$ directions, respectively, see Figure~\ref{fig:conecoords}a.

In conclusion, the change of coordinates \eqref{ccoorho}, implying \eqref{bendiag}-\eqref{defcone},  allows us to translate  \eqref{minrhouc} into the following problem
\begin{equation}
\min_{\gk\in\mathcal{C}}U(\gk),\qquad U(\gk):=\dfrac{\mathcal{E}_b}{2D}=\dfrac{1}{2} \Vert \gk-\bar\gk \Vert^2.
\label{mincone}
\end{equation}
This new formulation of the problem  depends on only two independent non-dimensional parameters, $a$ and $b$, which completely characterise the cone $\mathcal{C}$ of inextensible curvatures in terms of the material properties of the shell. It has also a remarkable geometric interpretation that will be largely exploited in the rest of this paper. We encapsulate this result  in the following proposition:
\begin{proposition}[Geometric analogy]
 After the change of coordinates~\eqref{ccoorho}, the admissible (uniform) curvatures of the {inextensible} Koiter shell are constrained to lie on the cone $\mathcal{C}$ defined by \eqref{conedef}. The potential  energy of the shell with (uniform) inelastic curvatures  $\bar\gk\in\real^3$ in a configuration $\gk\in\mathcal{C}$ is proportional to the squared Euclidean distance $ \Vert \gk-\bar\gk \Vert^2$. Hence,  given the inelastic curvatures  $\bar\gk\in\real^3$, the (locally) stable equilibria of the shell are all the points $\gk^{(i)}$  on the cone $\mathcal{C}$ having (locally) minimal Euclidean distance from the \emph{target} point $\bar\gk\in\real^3$. 
\end{proposition}
In an augmented Lagrangian approach, the stationarity condition of \eqref{mincone} can be written by imposing the gradient of the energy, $\partial U/\partial\gk=\gk-\bar\gk$, to be parallel to the gradient to the constraint $c(\gk)=0$ as defined in \eqref{conedef}:
\begin{equation}
\gk-\bar\gk=\lambda\, \frac{\partial{c(\gk})}{\partial \gk}, \qquad \frac{\partial{c(\gk})}{\partial \gk}=2\{-\kappa_m,\kappa_d/a^2, \kappa_t/b^2\},
\label{stationarity}
\end{equation}
where $\lambda$ is an additional scalar  unknown (Lagrange multiplier). This equation has a straightforward  interpretation in our geometric analogy that can be used to graphically solve the equilibrium problem: the equilibria are the points on the cone for which the  difference $\gk-\bar\gk$  is parallel to  ${\partial{c(\gk})}/{\partial \gk}$ and, hence, normal to the cone.
Moreover, being both the energy and the constraint quadratic functions of $\gk$, equation \eqref{stationarity} is linear and can be solved uniquely for $\gk$ as a function of $\lambda$. Replacing the solutions of \eqref{stationarity} into the constraint \eqref{defcone}, gives a quartic polynomial in $\lambda$ (the explicit calculations are not reported here). This proves that for any  inelastic curvature $\bar\gk$ there are at most four equilibrium configurations. 

In order to analytically describe the solutions of the equilibrium problem \eqref{stationarity}, we introduce the coordinates $(\kappa_m,\theta)$ on the cone such that any point on $\mathcal{C}$ is written as
\begin{equation}
\gk(\kappa_m,\theta)=\kappa_m \,\left(1,a \cos \theta, b \sin  \theta\right),
\label{konC}
\end{equation}
where $\kappa_m$ is the axial coordinate (we assume  $\kappa_m>0$) and $\theta\in[0,2\pi]$ is the angular coordinate along the cone cross-section (see Figure~\ref{fig:conecoords}a).
Hence, problem \eqref{mincone} can be recast as the unconstrained minimisation problem:
\begin{equation}
\min_{\kappa_m,\theta \in[0,2\pi]} \left\{U(\kappa_m,\theta)= \frac{(\kappa_m-\bar\kappa_m)^2}{2}+  \frac{(\kappa_m a \cos\theta-\bar\kappa_d)^2}{2}+  \frac{(\kappa_m b \sin\theta-\bar\kappa_t)^2}{2}\right\}
\label{mincone2}
\end{equation}
The stationarity conditions for this problem are $\partial U/\partial\kappa_m=0$ and $\partial U/\partial\theta=0$, which may be rewritten in the following form:
\begin{equation}
\kappa_m=\dfrac{\bar\kappa_m+a \,\bar\kappa_d \cos \theta +b \,\bar\kappa _t \sin \theta }{a^2 \cos ^2\theta +b^2 \sin ^2\theta +1}, 
\quad 
\bar\kappa_t\,\dfrac{b \left(1+a^2\right)}{a \left(1+b^2\right)}\cos\theta +  \bar\kappa_m\,\dfrac{a^2-b^2}{2 a\left(1+b^2\right)}\sin 2\theta= \bar\kappa_d\,\sin \theta.
\label{kmtheta}
\end{equation}
 Using the first of the expressions above, one can eliminate the axial variable $\kappa_m$ and rewrite the energy $U$ as a function of $\theta$ only, reducing the generic problem to the solution a one-degree-of-freedom system. The equilibria are the solutions $U'(\theta)=0$ and their stability will be given directly from the sign of $U''(\theta)$, where the prime denotes here the derivative with respect to $\theta$.\\
Given the value of the angular coordinate $\theta$ on the cone, the corresponding curvature of the shell is a uniaxial tensor $\k=k \,\bold{e}(\varphi)\otimes\bold{e}(\varphi)$
describing a cylindrical configuration, as depicted in Figure~\ref{fig:conecoords},  with   axis orientation $\varphi$ and  magnitude $k$ given by:
\begin{equation}
\varphi=\arctan\left(\frac{1}{\sqrt[4]{\beta}}\tan\left(\frac{\pi-\theta}{2}\right)\right)
,\qquad 
k = \kappa_m\frac{ 
1+\sqrt{\beta }+\cos \theta (1-\sqrt{\beta } )
   }{\sqrt[4]{\beta } \sqrt{\sqrt{\beta }+\nu }},
\label{phitheta}
\end{equation}
where  $\kappa_m$ is calculated from \eqref{kmtheta}.  The first equation above relates the angular coordinate $\theta$ on the cone with the  orientation $\varphi$ of the axis of maximal curvature of the actual shell configuration. For $\beta=1$, $\varphi=(\pi-\theta)/2$.
We recover below some  known literature results in our framework, giving useful insights for our design problem.

\subsection{Example: square-symmetric shell with isotropic inelastic curvature}
\label{sect:examples}

Let us consider the case of a shell made of a square-symmetric material with a given isotropic inelastic (\emph{target}) curvatures $\bar k_{11}=\bar k_{22}=\bar k$, $\bar k_{12}=0$.
A square-symmetric material is a material equally reinforced, or weakened, in two mutually orthogonal directions, and it is characterised by $\beta=1$ and $\alpha\neq 1$ (the material is isotropic for  $\alpha=1$). Then, the cone \eqref{conedef} has elliptical cross-sections with the semi-axes $a=\sqrt{(1-\nu)/(1+\nu)}$ and $b=\sqrt\alpha \,a$.
Moreover, the point $\bar\gk$ lies on the cone axis: according to the change of coordinates \eqref{ccoorho} and being $\beta=1$ for a square-symmetric material, one gets $\bar\kappa_d=\bar\kappa_t=0$, $\bar\kappa_m= \sqrt{1+\nu}\,\bar k$.
The equilibrium equation reduces to 
\begin{equation}
U'(\theta)=\frac{\bar\kappa_m^2 \left(b^2-a^2\right) \sin 2\theta  }{2\left(a^2 \cos ^2\theta +b^2
   \sin ^2\theta +1\right)^2}=0.
\label{eq:equilSqSym}
\end{equation} 

If the shell is not isotropic ($\alpha\neq1$,$a \neq b$), there are four solutions with $\theta^{(1,2,3,4)}=0, \pi/2, \pi, 3\pi/2$. By checking the sign of the second derivative of the energy, one can easily check that the stability of these solutions depends on $b/a=\sqrt\alpha$.  We conclude that the shell is bistable for any $\alpha\neq1$ and:
\begin{itemize}
 \item For $\alpha>1$,  the stable equilibria are $\theta^{(1,3)}=0,\pi$, corresponding to the two cylindrical configurations of curvature magnitude  $k=\bar k (1+\nu)$  oriented along the two material axes $\varphi= (\pi/2, 0)$. 
 \item For $\alpha<1$,  the stable equilibria are $\theta^{(2,4)}=\pi/2, 3\pi/2$, corresponding to the two cylindrical configurations of curvature magnitude  $k=2\bar k (1+\nu)/(1+\nu+\alpha-\alpha\nu)$
 oriented along the axes rotated by $\varphi=\pm \pi/4$ from the material ones. 
\end{itemize}
For a geometric illustration of these results on the cone,  see Figure~\ref{fig:coneisoaniso}.

For isotropic shells, $\alpha=\beta=1$, the cone of inextensible curvatures \eqref{conedef} has circular cross-sections, $a=b$. The equilibrium equation \eqref{eq:equilSqSym} degenerates, being trivially verified for any $\theta\in[0,2\pi]$. All the cylindrical configurations with a curvature magnitude $k=\bar k (1+\nu)$ are equilibria, independently of the orientation of the axis of maximal curvature.
 We say that the shell is \emph{neutrally stable} with the \emph{zero-stiffness} mode corresponding to the precession of the axis of maximal curvature (depicted by direction $\varphi$ in Figure~\ref{fig:conecoords}b). This case has been considered in \cite{GueKebPel11}, where the inelastic effects were imposed by plastic deformations of an isotropic metallic shell.
The results above become transparent from Figures~\ref{fig:coneisoaniso} bearing in mind the geometrical interpretation of the potential energy of the shell as the distance  $\Vert \gk -\bar\gk \Vert^2$.
All the  points lying on a circular cross-section of the cone have the same  minimal distance from the point $\bar\gk$, \emph{i.e.} the same elastic energy.

The result of the present inextensible shell model are valid in the "large curvature" regime. To make this approximation more precise, Figure~\ref{fig:compinext} compares the equilibrium curvature of the shell as a function of the isotropic inelastic curvature as obtained with the present inextensible model and an alternative uniform curvature model accounting for membrane deformations \cite{SefMcM07}. In the extensible model the equilibrium shape is unique up to a critical inelastic curvature 
\begin{equation}
\bar k^* = \frac{8}{(1+\nu)^{3/2}}\frac{h}{L^2}, \quad\text{or}\quad \bar{\kappa}_m^*=\frac{8}{1+\nu}\frac{h}{L^2},
\label{eq:bifextensible}
\end{equation}
where $h$ and $L$ are the thickness and the radius of the initially flat disk, respectively.
The extensible model predicts multiple stable shapes for  $\bar k> \bar k^*$  and its results tend to those of the present approximate inextensible model for $\bar k\gg \bar k^* {h}/{L^2}$ (see {\cite{SefMcM07}} or \cite{Ham16} for the detailed calculations). Only the case of an isotropic material ($\alpha=1$) is reported here, but analogous considerations are valid for a generic orthotropic shell. As shown in \cite{SefMcM07,SefMau13}, the results of the extensible uniform curvature model are very close to those of full-field, fully nonlinear,  finite element calculations.

{\begin{figure}
\centering
{
\subfloat[$\alpha <1$]{\includegraphics[width=.35\textwidth]{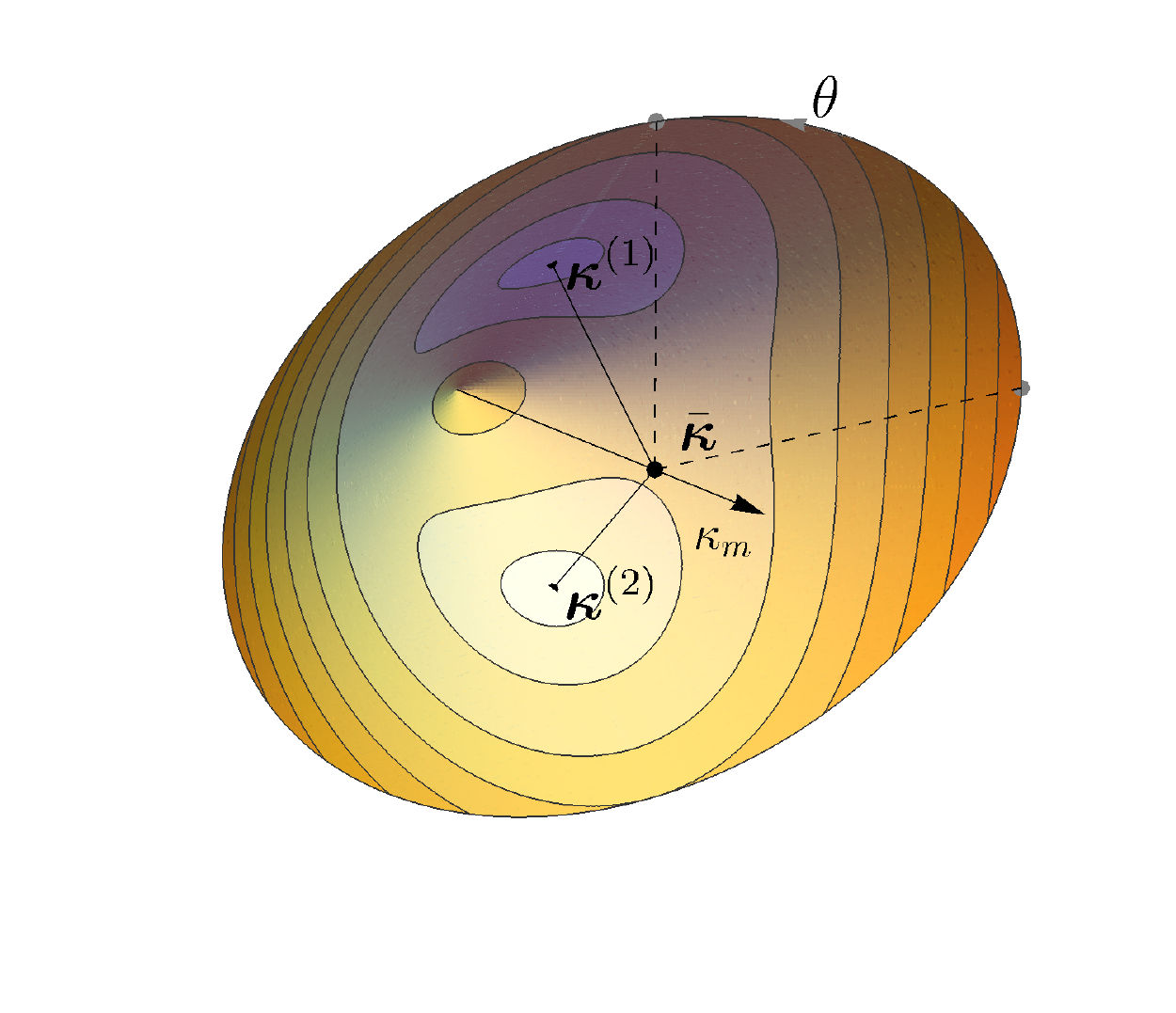}}
\subfloat[$\alpha =1$]{\includegraphics[width=.35\textwidth]{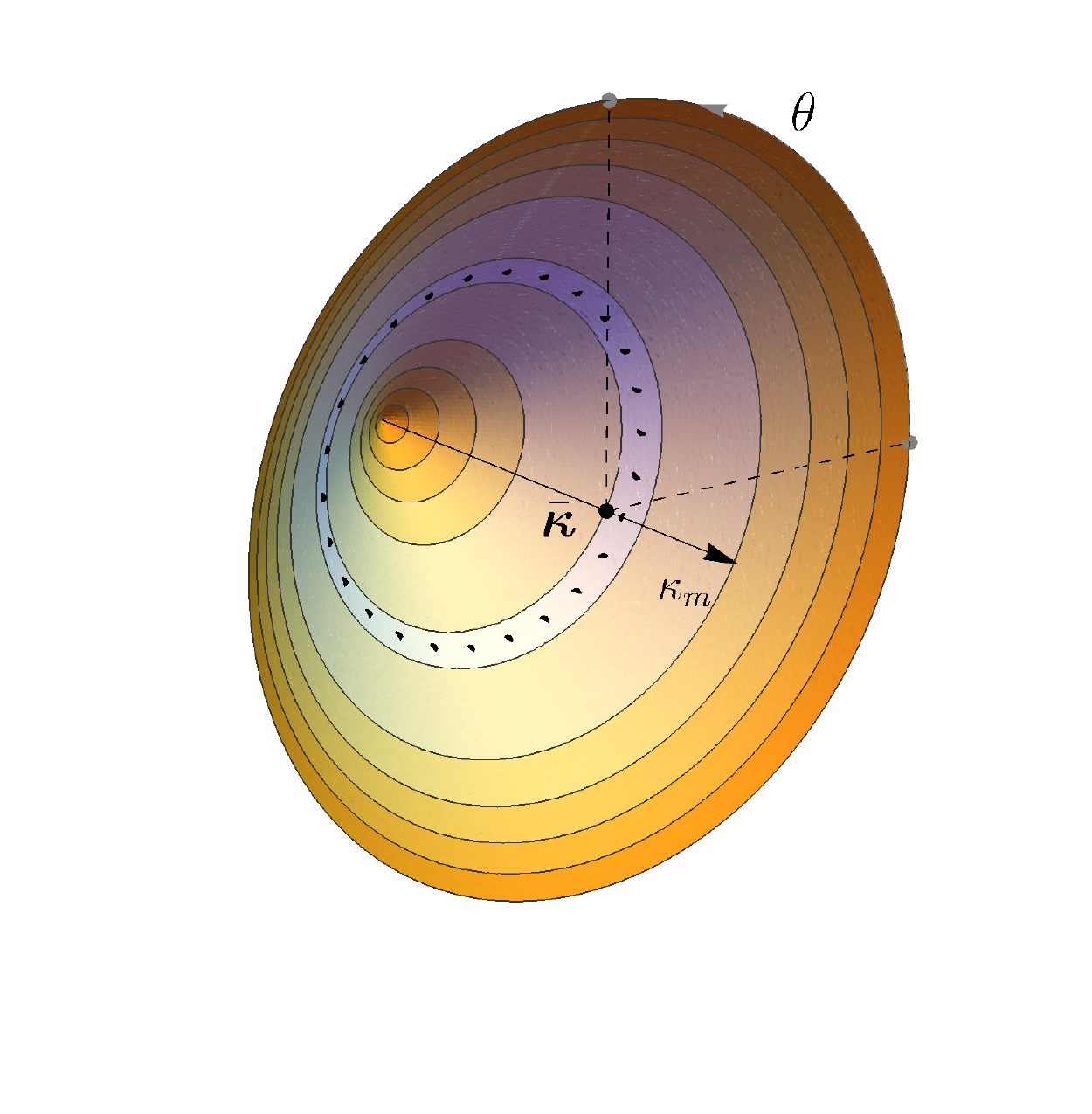}}
\subfloat[$\alpha >1$]{\includegraphics[width=.35\textwidth]{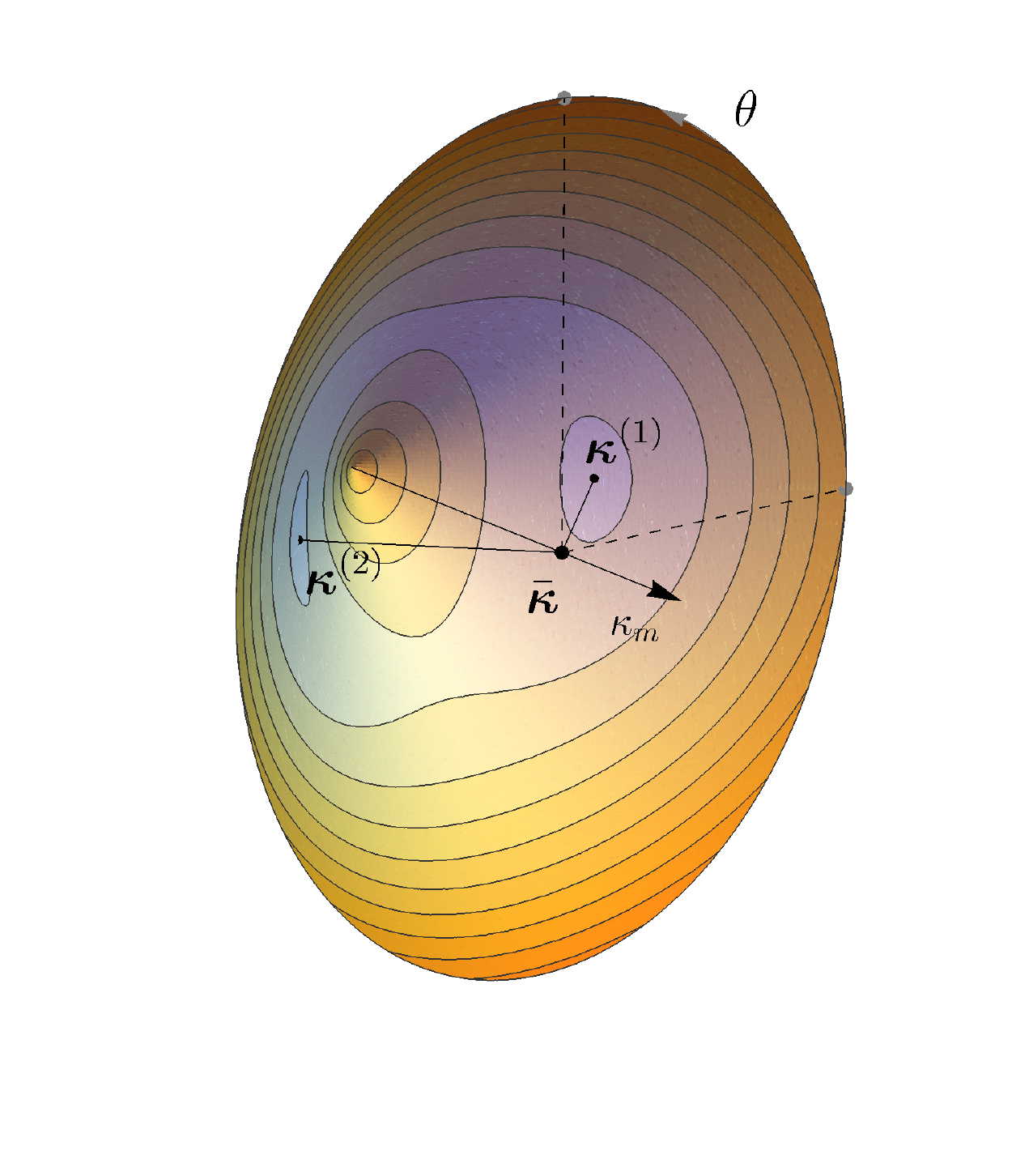}}
}

\caption{Square-symmetric shells with isotropic inelastic curvatures: $\beta=1$, $\bar k_{11}=\bar k_{22}=\bar k$, and $\bar k_{12}=0$.
(a), (b) and (c) show the  contour plot of the potential energy \eqref{mincone2} on the cone representing the inextensible constraint in the $(\kappa_m,\kappa_d,\kappa_t)$ space (minimal values are in lighter color). For $\alpha<1$ (a) and $\alpha>1$ (c) the shell is bistable, whilst in the isotropic case $\alpha=1$ (b), the shell is neutrally stable. The energy barrier to pass from $\gk^{(1)}$ to  $\gk^{(2)}$ is proportional to $\vert\alpha-1\vert$.  
}
\label{fig:coneisoaniso}
\end{figure}

\begin{figure}
\centering
\subfloat[]{\includegraphics[width=.43\textwidth]{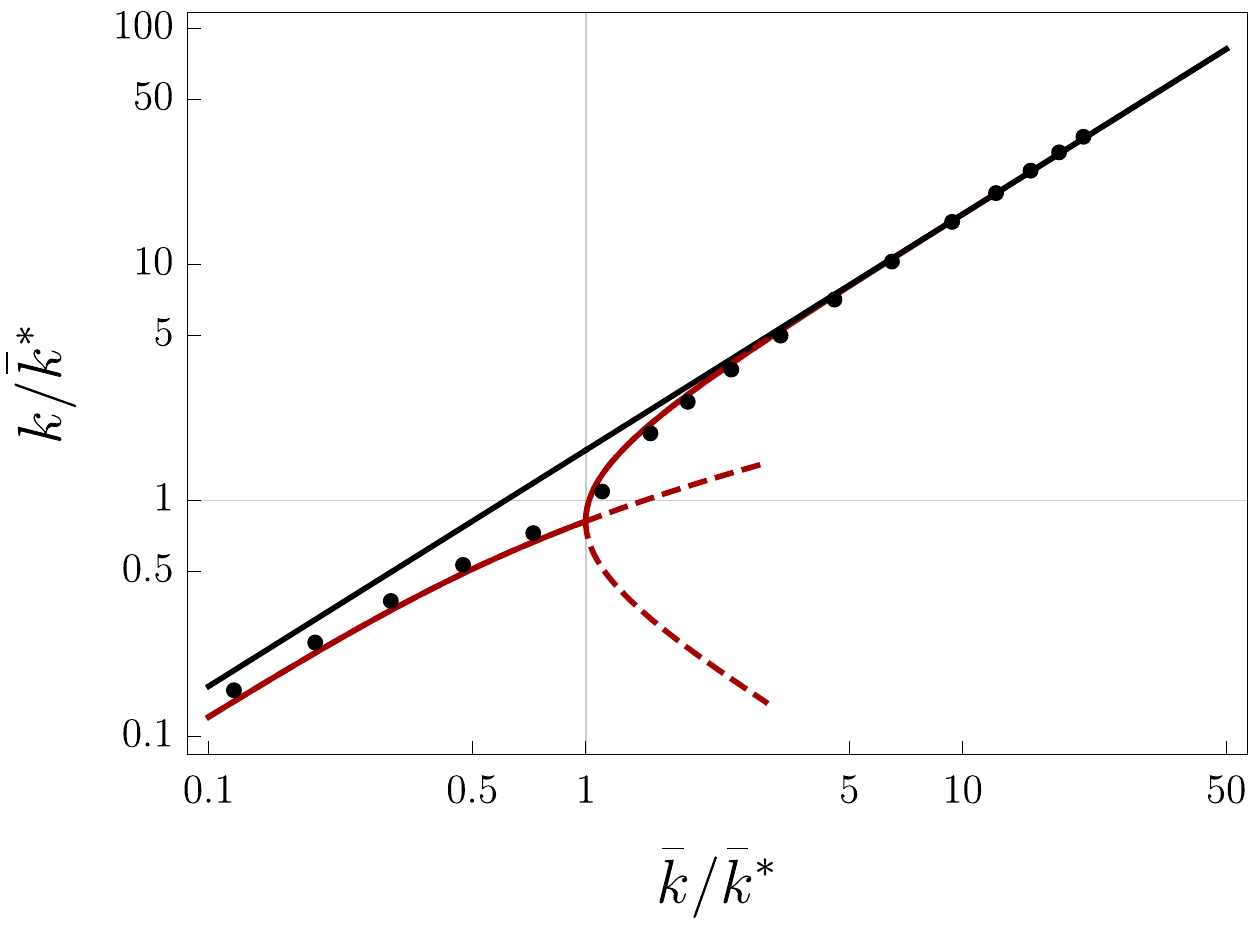}}\qquad
\subfloat[]{\includegraphics[height=.32 \textwidth]{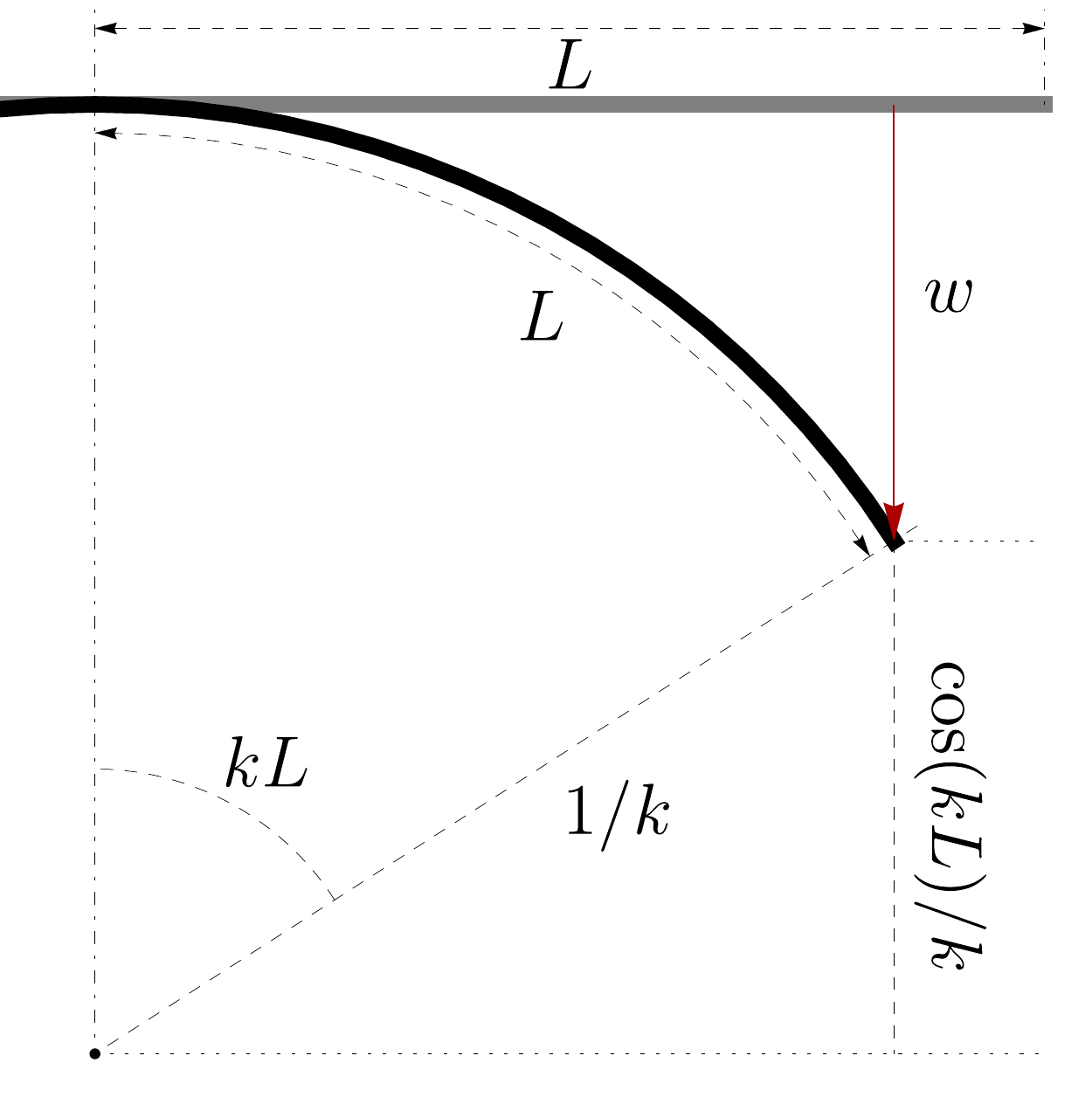}}
\caption{(a): Comparison of the equilibrium curvatures predicted by the extensible (red) \cite{SefMcM07} and inextensible (black) uniform curvature models, and the finite element simulations (dots). The present inextensible model gives a very good approximation of the inextensible model as soon as $\bar k>2 \bar k^*$, see  equation \eqref{eq:bifextensible}. (b): Schematic illustration of the basic trigonometric identities used to obtain the relation \eqref{wkrel} between the (uniform) curvature $k$ and the maximum peripheral displacement $w$.
}
\label{fig:compinext}
\end{figure}

\section{Design of the actuation law}\label{sect:act}
We consider the problem of designing  an actuation strategy to control the  shape of the shell by applying inelastic curvatures of piezoelectric origin. 
In practice, a fundamental constraint is the maximal inelastic curvature that the embedded actuation can exert. 
In a structure with a vanishing stiffness deformation mode, we design an actuation law that moves the configuration of the shell along the direction of minimal stiffness at each instant. 
This strategy will reduce the actuation energy requirements, but it will demand a suitable multiparameter actuation.

We focus on the case of the cylindrical shell obtained by applying isotropic inelastic curvatures to an initially flat disk, which has been treated in the previous Section. The geometric analogy of the cone facilitates the design of suitable actuation strategies in terms of inelastic curvatures. 
Our goal is to move the configuration of the shell by rotating the curvature axis along the (almost) zero-stiffness mode corresponding to ellipses of (almost) constant energy on the cone, as illustrated in Figure~\ref{fig:coneisoaniso}. These ellipses are parametrically described by a curvature tensor $\gk(t)$ in the form
\begin{equation}
\gk(t)=\kappa_m\{1, a \cos t, b \sin t\}, t\in [0,2\pi]\}.
\label{shapecontrol}
\end{equation}   
In order to drive the \emph{precession} of the shell according to \eqref{shapecontrol}, the basic idea is to apply inelastic curvatures $\bar\gk(t)$ that describe ellipses centered on the cone axis at coordinate $\bar\kappa_m$ as follows 
\begin{equation}
{\bar\gk}(t)= \{\bar\kappa_m, 0,0\}+\{0,\,\bar\rho \,a\, \cos t, \,\bar\rho\,b\, \sin t\}, \quad \bar\kappa_m > 0, \quad\bar\rho \ge 0.
\label{actuationlaw}
\end{equation}
The first contribution stands for the isotropic plastic curvature of amplitude $\bar\kappa_m$. The second one is a perturbation of amplitude $\bar\rho$ and phase $t$, controlled by the piezoelectric actuators. In practice, because of the limited actuation strains provided by piezoelectric materials, the amplitude of the  piezoelectric actuation $\bar\rho$ should be considered much smaller of the plastic curvature $\bar\kappa_m$. In turn, $\bar\kappa_m$ should be sufficiently large to put the initially flat disk within the geometric non-linear regime, namely  $\bar\kappa_m\gg \bar{\kappa}_m^*$, see \eqref{eq:bifextensible}. 

If the shell is perfectly isotropic ($\alpha=1,a=b$), a perturbation with an arbitrarily small amplitude $\bar \rho$ leads to a sequence of equilibrium points on the cone close to the desired circular path \eqref{shapecontrol}. This  physically means that, in a perfect isotropic system, a vanishing actuation ($\bar \rho\to 0$) is sufficient to drive a continuous precession of the maximal curvature axis along the set of neutrally-stable configurations.

In practice, the conditions for perfect isotropy are never realised and the real behaviour of the structure is different.  We   show that a minimal actuation threshold is required to drive the precession of the shell, and we  relate this threshold to the imperfections. To this end, we consider the actuation problem for the  more realistic case of weakly anisotropic shells with $\alpha\neq 1$, for which the cone forming the inextensible curvature manifold has an elliptic cross-section with $a\neq b$, see \eqref{defsemiaxes} and Figure~\ref{fig:coneisoaniso}.

After introducing \eqref{actuationlaw} into the second equation of \eqref{kmtheta}, and some basic manipulations, we find that the equilibria, in the form \eqref{konC}, should satisfy 
\begin{equation}
(1+\tan^2\theta)\left( \tan\theta-\dfrac{b^2(1+a^2)}{a^2 (1+b^2)} \tan t\right)^2 =\left (\dfrac{a^2-b^2}{a^2 (1+b^2)} \, \dfrac{\bar\kappa_m }{\bar\rho\, \cos t}\right)^2 \tan^2\theta.
\label{theta2}
\end{equation}
which, being a fourth order polynomial in $\tan\theta$, admits at most four distinct solutions. We study how these solutions $\theta^{(i)}(t)$ of \eqref{theta2} depend on the amplitude of the actuation $\bar\rho$ when $t$ varies from $0$ to $\pi$ (the behaviour in $[\pi,2\pi]$ will be symmetric). We focus on the case with $\alpha>1$ ($b>a$), for which at $t=0$ the solution $\theta=0$ of  \eqref{theta2} is stable, as shown in the previous Section and Figure~\ref{fig:coneisoaniso}c. The case with $\alpha<1$ can be treated with minor modifications. 
Remember that, for $\beta=1$, an angle $\theta$ on the cone corresponds to an orientation $\varphi=(\pi-\theta)/2$ of the maximal curvature axis of the equilibrium configuration, see \eqref{phitheta}.  
\begin{figure}[h]
\centering
\begin{tabular}{ccc}
$ \bar\rho > \bar\rho^{**}$ &$\bar\rho^{*} < \bar\rho < \bar\rho^{**}$ &$\bar\rho<\bar\rho^{*}$  \\
\includegraphics[width=.26\textwidth]{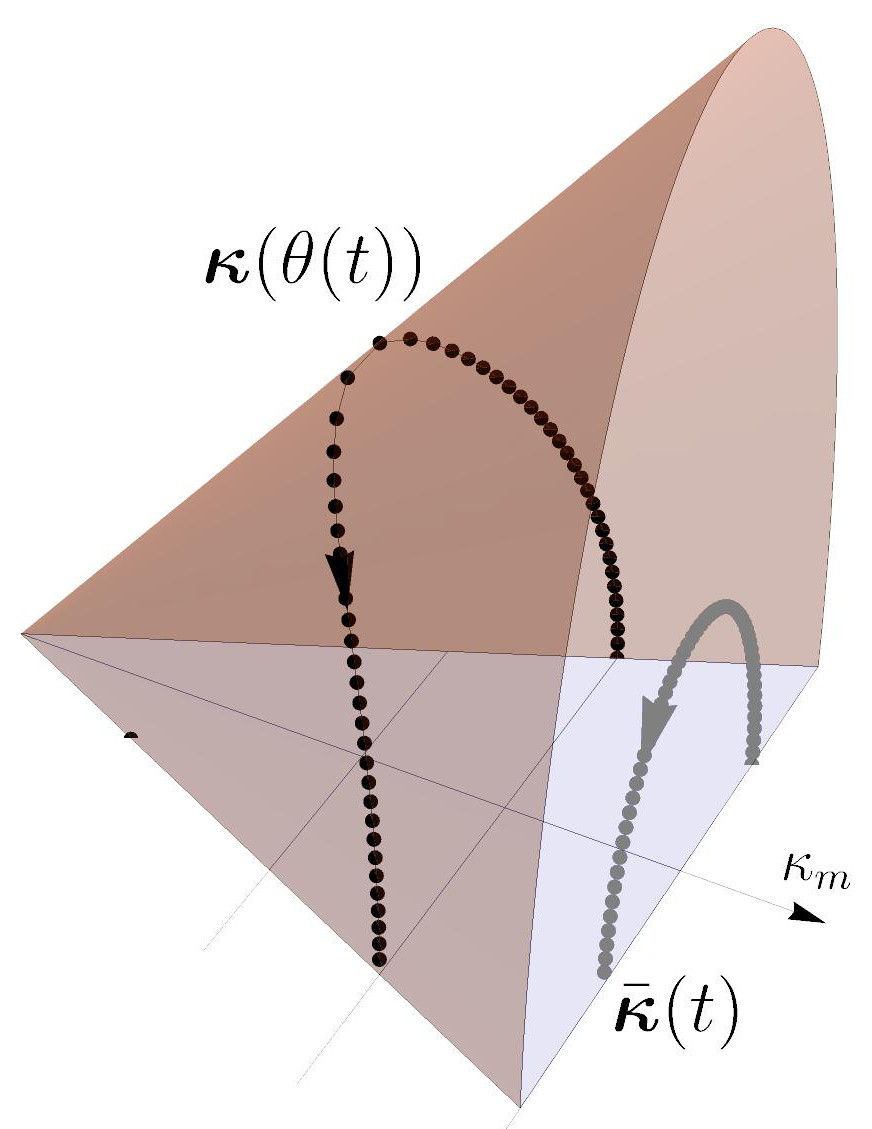}&
\includegraphics[width=.26\textwidth]{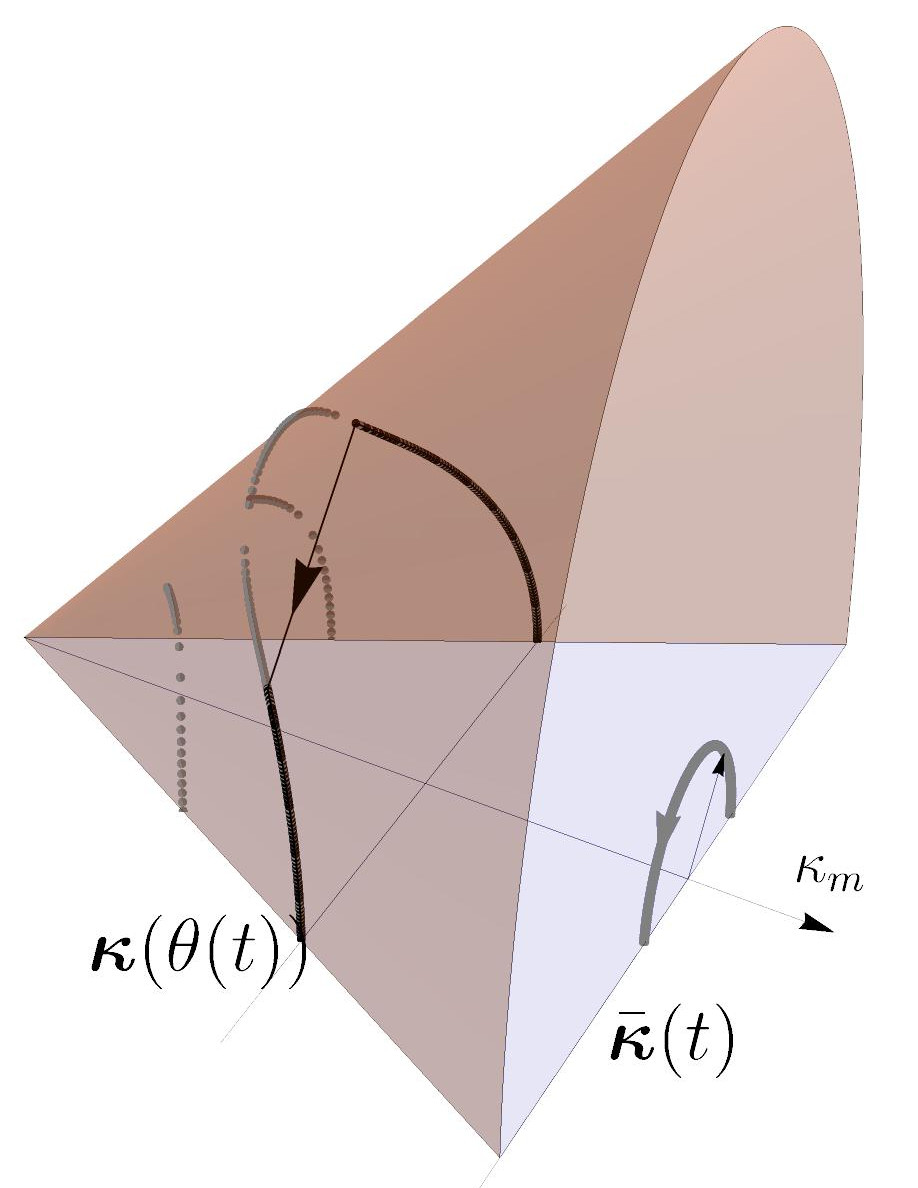}&
\includegraphics[width=.26\textwidth]{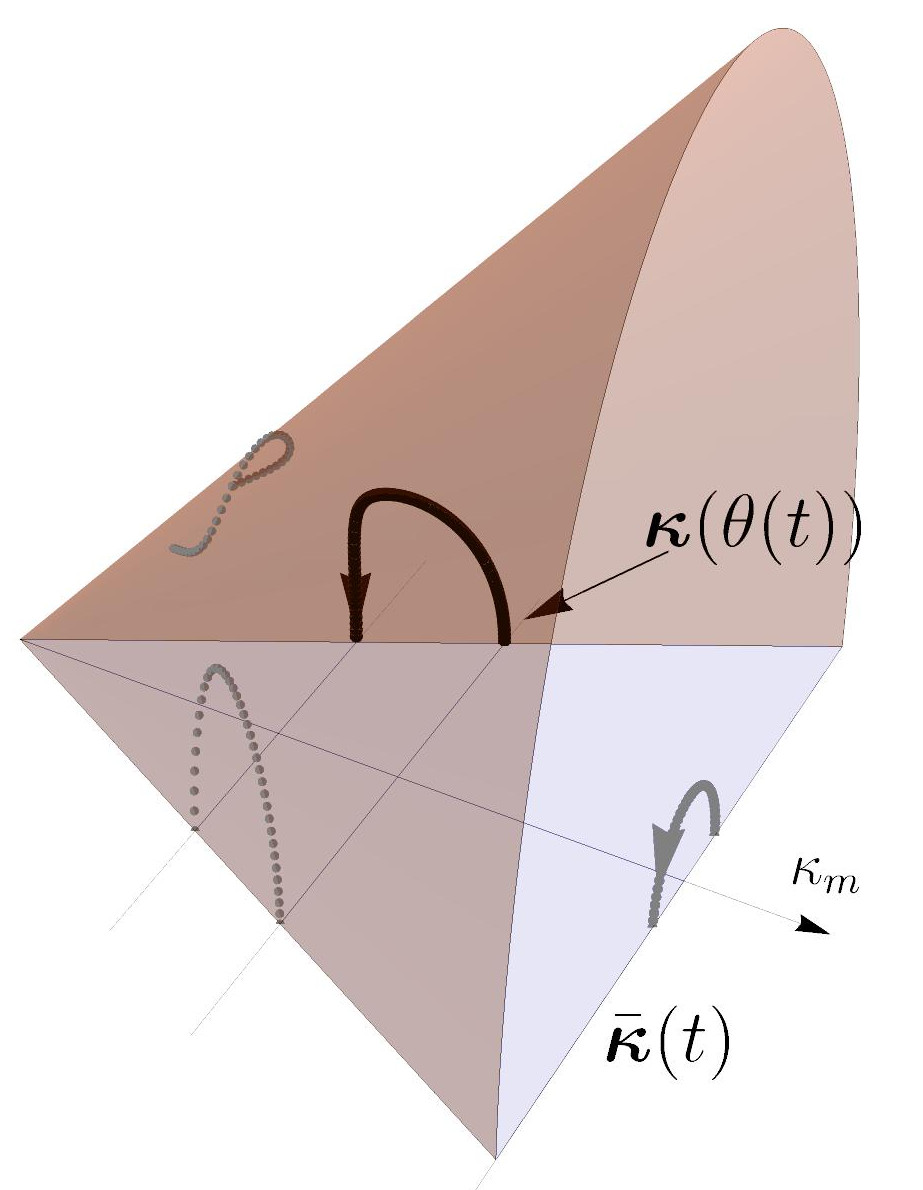}\\
 \includegraphics[width=.3\textwidth]{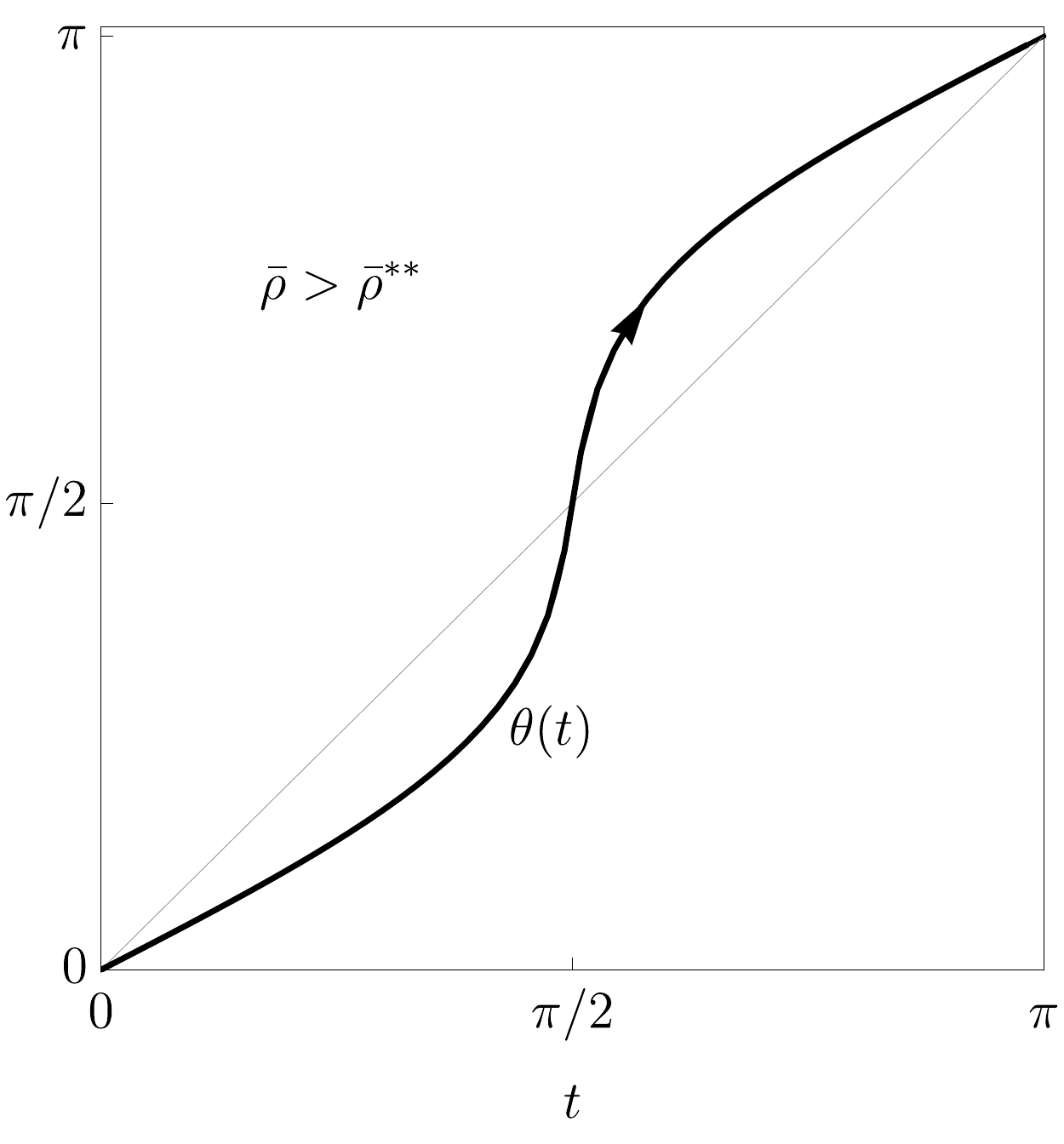}&
\includegraphics[width=.3\textwidth]{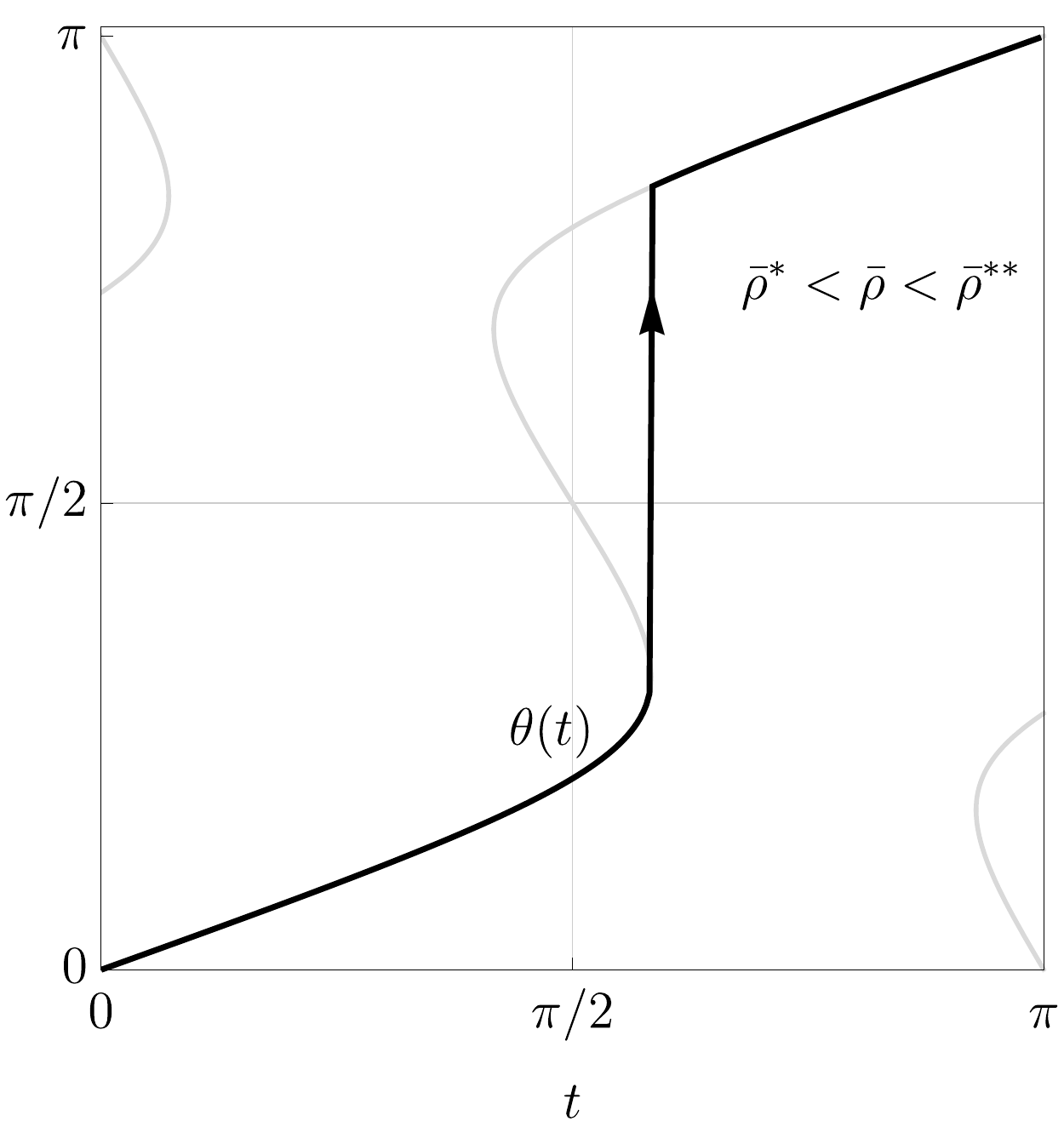}&
\includegraphics[width=.3\textwidth]{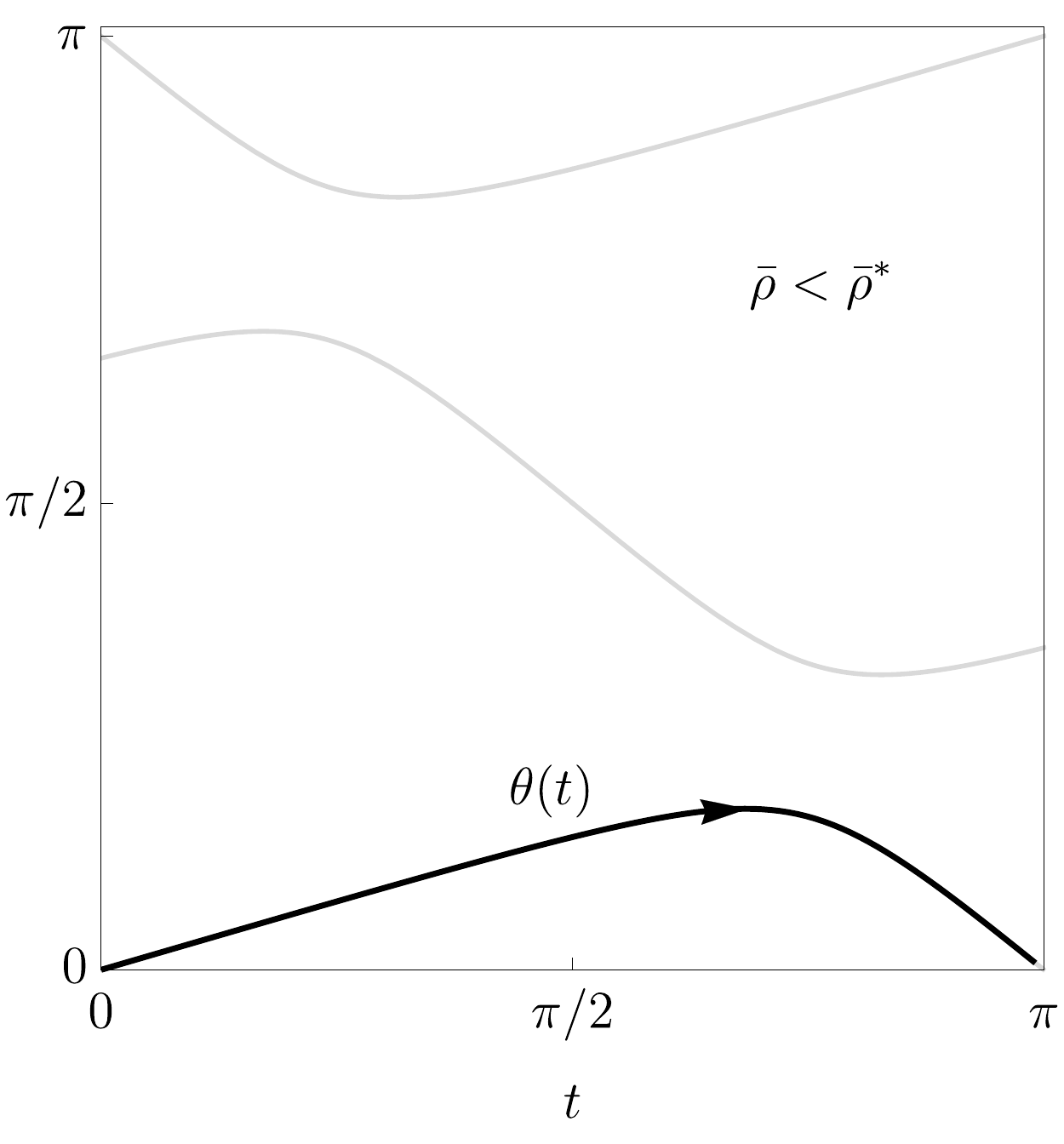}\\
 (a) & (b)   & (c)  \\
\end{tabular}
\caption{In columns, the three possible actuation regimes according to the relative value of the actuation amplitude $\bar\rho$ with respect to the thresholds in  \eqref{rhostar}. The figures in the top row show the loci of the inelastic curvature $\gh(t)$ (grey) and of the resulting curvature at the equilibrium $\gk(\theta(t))$ (black) on the cone $\mathcal{C}$. The bottom row reports the correspondence between the actuation phase $t$ and the angular coordinate $\theta$; the gray curves indicate (stable or unstable) equilibria, the black curves indicate the actual paths followed by the shell configuration.}
\label{fig:cases}
\end{figure}

It turns out that there are two thresholds for the actuation amplitude     $\bar\rho$, namely
\begin{equation}
\bar\rho^*:= \dfrac{\bar \kappa_m\,(b^2-a^2)}{a^2+b^2+2 a^2 b^2},\qquad \bar\rho^{**}:=\dfrac{\bar \kappa_m\,(b^2-a^2)}{b^2+a^2 b^2}\ge\bar\rho^{*},
\label{rhostar}
\end{equation} 
separating the behaviour of the shell in three different regimes, as sketched in Figure~\ref{fig:cases}:
\begin{enumerate}
\item for $\bar\rho>\bar\rho^{**}$ there is only one stable branch defining a continuous and monotonic relation between the actuation phase $t\in [0,\pi]$ and the angular coordinate $\theta\in [0,\pi]$ on the cone. This stable branch, $\theta=\theta^{(1)}(t)$, stems from $(t=0,\theta=0)$ and arrives for $t=\pi$ in $\theta=\pi$; it is drawn as a thick black curve in Figure~\ref{fig:cases}a for $\bar\rho/\bar\kappa_m=0.5$, $a=1$, $b=1.5$. A similar actuation is therefore sufficient to drive continuously the stable equilibrium around the cone. Correspondingly, the maximal curvature axis of the  shell configuration  undergoes a complete quasi-static precession.

\item for $\bar\rho^{**}>\bar\rho>\bar\rho^{*}$ there are several branches; the branch stemming from $\theta=0$ for $ t=0$ is stable until it reaches a turning point for some $t=t_c>\pi/2$ and $\theta=\theta_c<\pi/2$. Thus for $t>t_c$ the equilibrium must necessarily snap-through towards another stable branch, namely the one passing in $\theta=\pi$ for $t=\pi$, see Figure~\ref{fig:cases}b. A similar actuation is sufficient to drive the shell equilibrium around the cone but not continuously. At $t=t_c$ the shell will dynamically snap with a sudden jump in the orientation of the maximal curvature axis.

\item for $0<\bar \rho<\bar \rho^*$ again there are several branches;  
the branch $\theta^{(1)}(t)$ stemming from $\theta=0$ for $t=0$ is stable and continuous but along this branch the angular coordinate $\theta$ remains strictly lower than $\pi/2$ for every $t\in[0,\pi]$. Moreover, for $t\to \pi$ we have $\theta^{(1)}(t)\to 0$, see Figure~\ref{fig:cases}c. Such level of actuation is not sufficient to drive the stable equilibrium of the shell around the cone.  
\end{enumerate}

In conclusion, the actuation law \eqref{actuationlaw} with $t \in[0,2\pi]$ is able to drive  a complete precession of the maximal curvature axis, if the actuation radius $\bar\rho$ is larger than the threshold value $\bar\rho^{*}$. The transition is quasi-static if $\bar\rho>\bar\rho^{**}$, but is dynamic otherwise. 
The two thresholds are bounded  $0\le \bar \rho^*\le\bar \rho^{**} \le \bar\kappa_m\,(b-a)$.
Hence, as the level of anisotropy vanishes  ($\alpha\to1$, $\beta\to 1$), we have $b\to a$ and both $\bar\rho^{*} \to 0$ and $\bar\rho^{**}\to 0$. As anticipated, even a small  actuation radius would then be sufficient. Moreover, in the limit $b\to a$, Eq. \eqref{theta2} reduces to $\tan\theta=\tan t$; thus as isotropic conditions are approached, the actuation angle and the angular coordinate on the cone tend to coincide.

\section{Experimental prototype and results}\label{sect:exp}
\label{sec:exp}
We develop an experimental prototype implementing the ideas of  the previous Section.
We obtain a bistable shell by plastically deforming an initially flat copper disk through a large inelastic isotropic curvature $\bar{k}_{\textsc{P}}$. 
Special care has been devoted to achieve almost isotropic conditions ($\alpha\simeq 1$), thus obtaining an almost neutrally-stable shell and reducing the energetic gap between the minima. Hence, we design a multi-parametric piezoelectric actuation to introduce a controlled inelastic curvature of piezoelectric origin, $\bar{\mathbf{k}}_{\textsc{V}}(t)$, so that the overall inelastic curvature, 
\begin{equation}
\bar{\mathbf{k}}(t)=\bar{k}_{\textsc{P}}\,\left(
\begin{array}{c}
1 \\1\\ 0
\end{array}
\right)+
 \bar{\mathbf{k}}_{\textsc{V}}(t),
\label{actuationsplitk}
\end{equation}
will vary as in \eqref{actuationlaw} and Figure~\ref{fig:cases}.
To this end, we use  three pairs of piezoelectric patches bonded on the upper surface of the shell, as shown in Figures~\ref{fig:configurationpzt}a-b, and oriented along the directions $ {\varphi}_1=\pi/2$, $ {\varphi}_2=-\pi/6$, $ {\varphi}_3=\pi/6$. 
This configuration is  inspired by strain-gauge rosettes used in classical  devices for tensorial strain measurement. It is designed to preserve the global isotropy of the disk, when considering uniform curvature deformation modes.  

\subsection{Plastic forming process}
\label{sec:plastification}
We start with a thin copper disk of circular shape (radius $L=122.5\,\text{mm}$ and thickness $h=0.3\,\text{mm}$). The disk is, to a good approximation, initially stress-free and flat.
It is made of a nominally  isotropic copper (\texttt{Cua1 H14}) with Young modulus $E\simeq 124\,$GPa, Poisson ratio $\nu\simeq 0.33$, and mass density $\delta \simeq 9690$ Kg/m$^3$. The flat disk has been manufactured through a rolling process. With this geometry and material properties, the critical inelastic curvature \eqref{eq:bifextensible}  for entering in the multistable regime is calculated to be
$\bar k^*\simeq0.1\,\mathrm{m}^{-1}$.

We apply an almost isotropic  plastic curvature by manually winding the copper disk in two orthogonal directions around PVC cylinders, see Figure~\ref{fig:pla}a. In order to achieve a homogeneous state of plastic deformation all over the structure,
several cycles of plastification are repeated with PVC cylinders of progressively smaller diameters ($160\,\text{mm}$, $140\,\text{mm}$, and finally  $110\,\text{mm}$). 
Because of the initial manufacturing process of the base material (rolling), the  plastic yielding properties are not perfectly isotropic. They show a preferred material orientation aligned with the rolling direction. 
The results reported below are obtained by aligning one of the winding directions, that we will denote by $\mathbf{e}_1$ and  $\mathbf{e}_2$, with the rolling direction of the copper disk ($\mathbf{e}_1$).
At the end of the forming process the disk shows two stable, almost cylindrical, configurations. 
Excluding some narrow zones near the edges, the shell curvature at the equilibria is almost uniform.
The axis of maximum curvature is oriented either with the direction $\mathbf{e}_1$ (configuration {(a)}, Figure~\ref{fig:configurationpzt}a, with $\varphi=0$) or $\mathbf{e}_2$  (configuration (b), Figure~\ref{fig:configurationpzt}b, with $\varphi=\pi/2$).  We estimate the curvature of the two configurations by measuring the maximal transverse displacement $w$ at the peripheral points through the following simple geometrical relation (see Figure~\ref{fig:compinext}b): 
\begin{equation}
w=\dfrac{1-\cos(k\,L)}{k}=\dfrac{k\;L^2}2+o(k^2).
\label{wkrel}
\end{equation}
The measured displacements are presented in the first row of Table~\ref{tab:identification}.
They correspond to a curvature $k^{(a)}\simeq 5.7±\pm0.1\,\mathrm{m}^{-1}$ for the first configuration ($\varphi=0$) and $k^{(b)}\simeq\,5.5±\pm0.1\,\mathrm{m}^{-1} $ for the second one ($\varphi=\pi/2$). Being  $k/\bar k^*\sim 50$, the use of an inextensible model is fully justified, see Figure~{\ref{fig:compinext}a}. On the basis of the results of Section~\ref{sect:0stiff}\ref{sect:examples},  the bistability and the kind of curvature at the equilibria lead us to infer that the shell is not perfectly isotropic. The observed behaviour is consistent to the one predicted for a square-symmetric shell, namely $\beta=1$ and $\alpha>1$. For this case, the relation between the maximal curvature at the equilibrium and the magnitude of the isotropic inelastic curvature is $k=\bar k (1+\nu)$. Hence,  we can estimate the inelastic curvature induced by the plastification process as $\bar k_\textsc{P}=k/ (1+\nu)=4.2 \pm0.1\,\mathrm{m}^{-1}$ taking $k$ as the average of the curvatures $k^{(a)}$ and $k^{(b)}$ of the two configurations.
 
Assuming $\beta=1$, we perform a  dedicated experiment to identify an appropriate value of the dimensionless shear stiffness parameter $\alpha$. We expect this parameter to be different from unity because of the observed bistable behaviour of the shell and we speculate that this can be a consequence of an anisotropic hardening during the plastification process. With the shell clamped at its center and initially in the configuration (a), we hang two identical  masses at the points indicated in the inset of Figure~\ref{fig:identification}a.  
We detect a critical value of the mass for which the equilibrium configuration  (a) with $\varphi=0$ loses its stability and the shell switches to the configuration (b) with $\varphi=\pi/2$. As detailed in the Appendix, this critical mass can be easily related to the dimensionless shear stiffness parameter $\alpha$ in the uniform curvature inextensible model. We repeated the analog experiment starting from the configuration (b). We find critical masses  $m_c\simeq 32\pm 5\text{g}$. In particular we observed an important variation of the value of the critical mass when leaving the shell in a given configuration for several hours or days. We attribute this effect to viscoelastic relaxation phenomena in the prestressed shell, see for instance \cite{Gigliotti2014}. Manually alternating several times the configuration of shell between the two stable equilibria can reduce this effect, restoring the initial almost isotropic pre-stress distribution. We adopted this expedient as a good practice before performing quantitative experiments. Using the relation \eqref{alphaid} of the Appendix, we found $\alpha\simeq 1.08\pm0.02$. The results are presented in the first row of Table~\ref{tab:identification}. Finally, after the plastification, the parameters of the uniform curvature model \eqref{minrhouc} are
\begin{equation}
D = 0.0148\, \mathrm{N\,m^3}, \quad \beta=1,\quad  \nu = 0.33,\quad \alpha = 1.08, \quad \bar{k}_\textsc{P}=4.2\mathrm{m}^{-1}.
\label{CPLTdisk}
\end{equation}

\begin{figure}[h!]
\centering
\begin{tabular}{cc} 
\includegraphics[height=.25\textwidth]{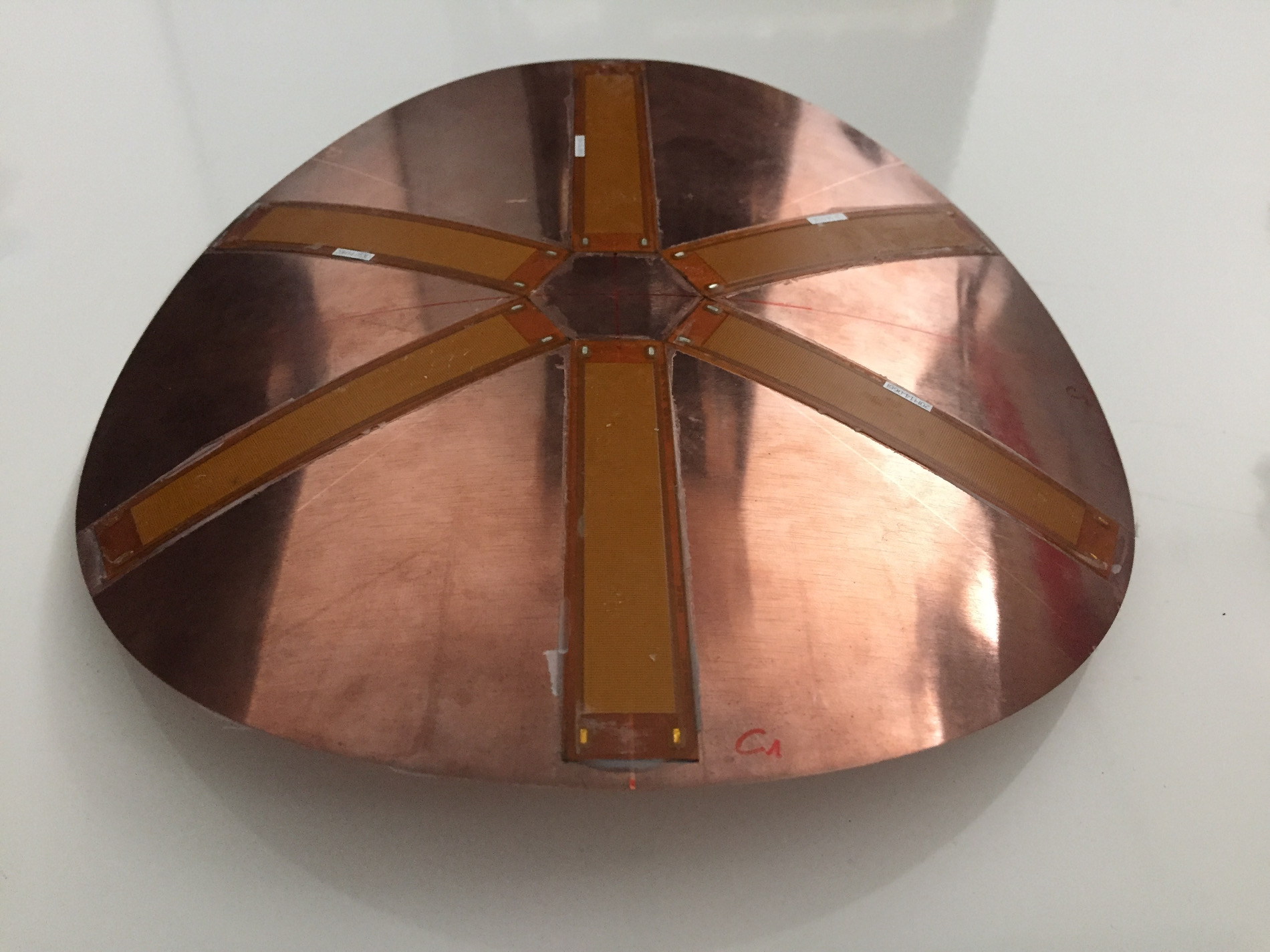} &
\includegraphics[height=.25\textwidth]{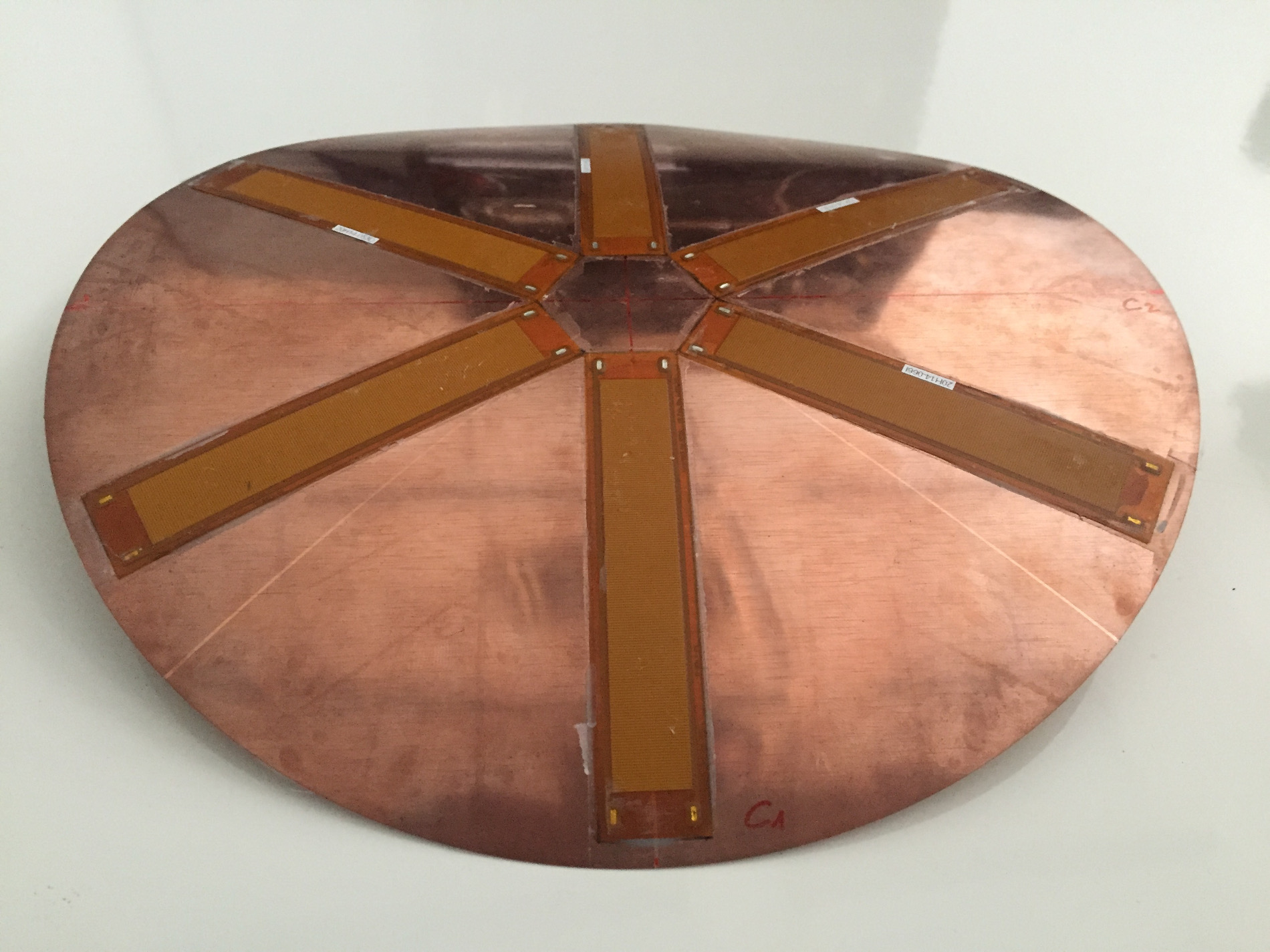}  \\
(a) & (b) 
\end{tabular}
\caption{ The two stable configurations and of the plastically deformed disk with the MFC actuators, with $\varphi=0$ (a) and $\varphi=\pi/2$ (b), respectively. }
\label{fig:configurationpzt}
\end{figure}

\begin{figure}[h!]
\centering
\begin{tabular}{cc} 
\includegraphics[width=.45\textwidth]{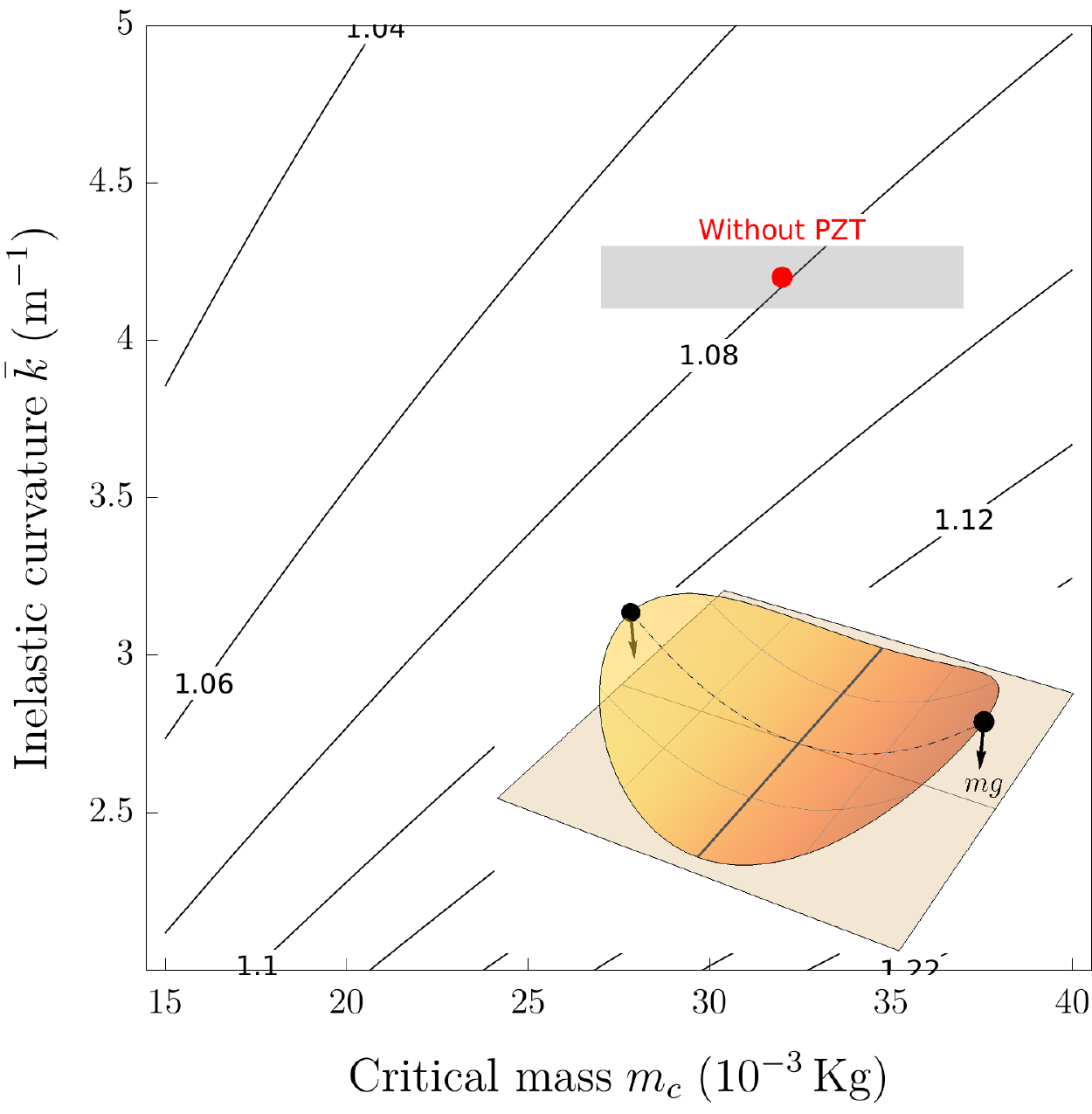}&
\includegraphics[width=.45\textwidth]{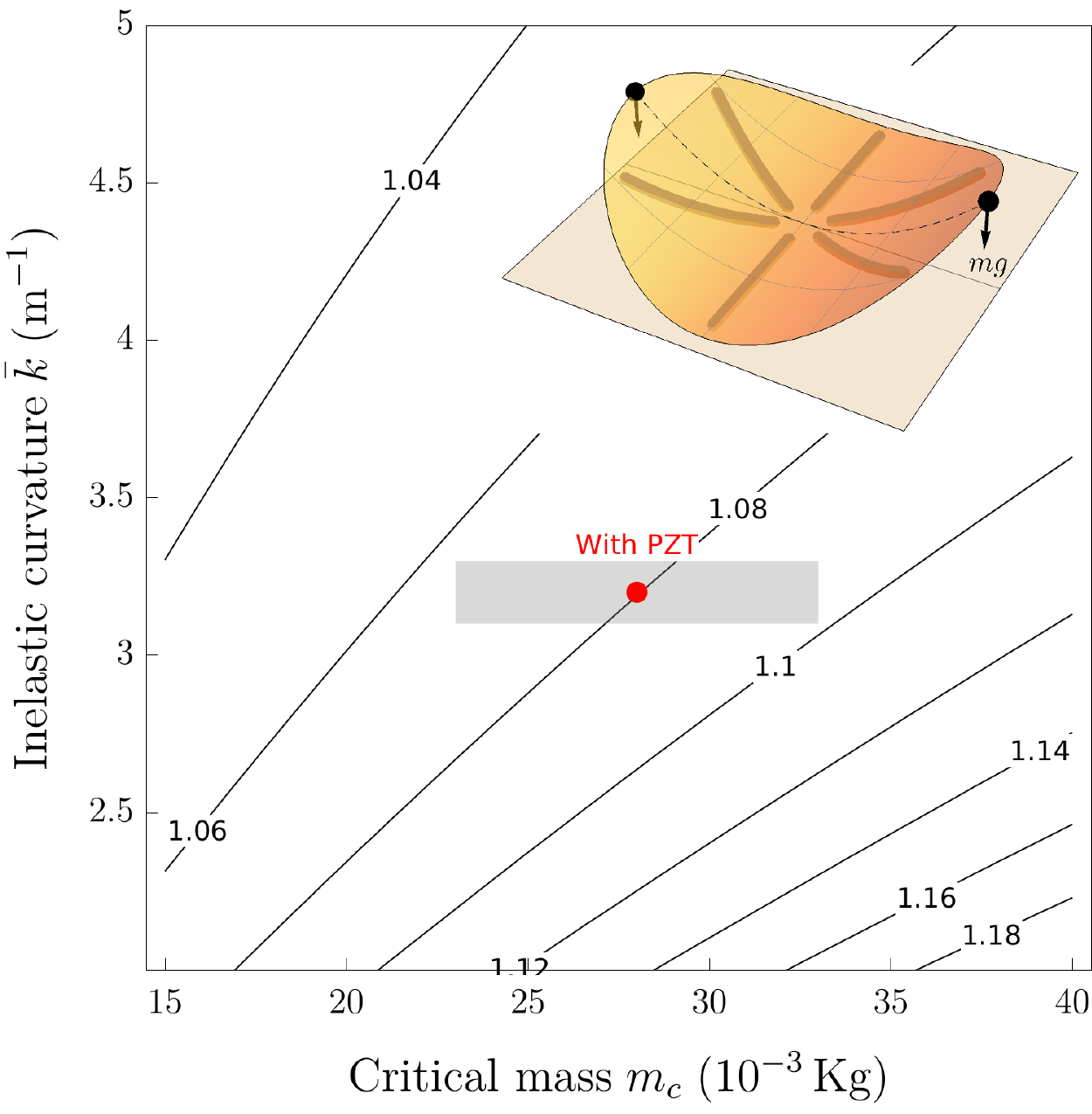}\\
(a) & (b)  \\
\includegraphics[width=.45\textwidth]{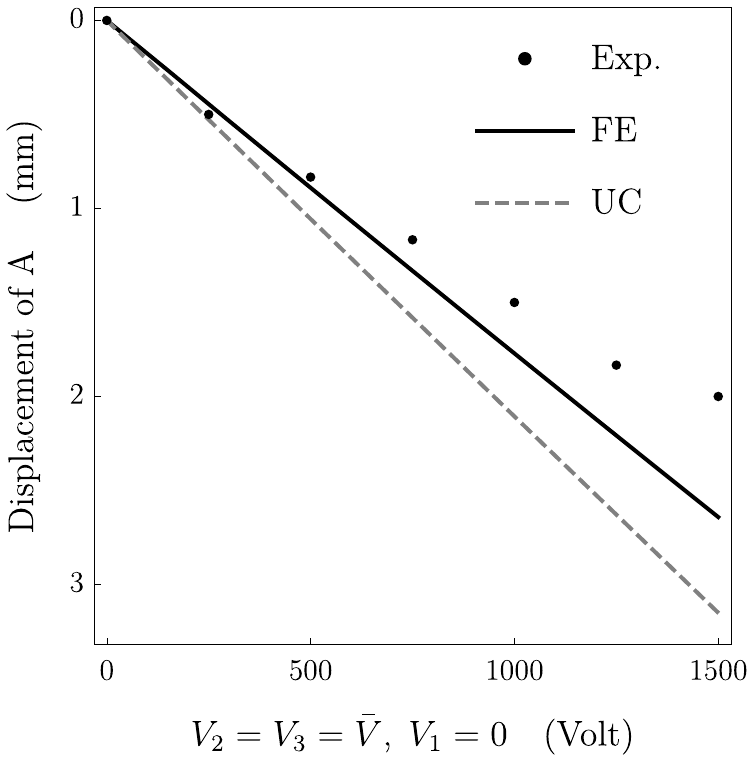}&
\includegraphics[width=.35\textwidth]{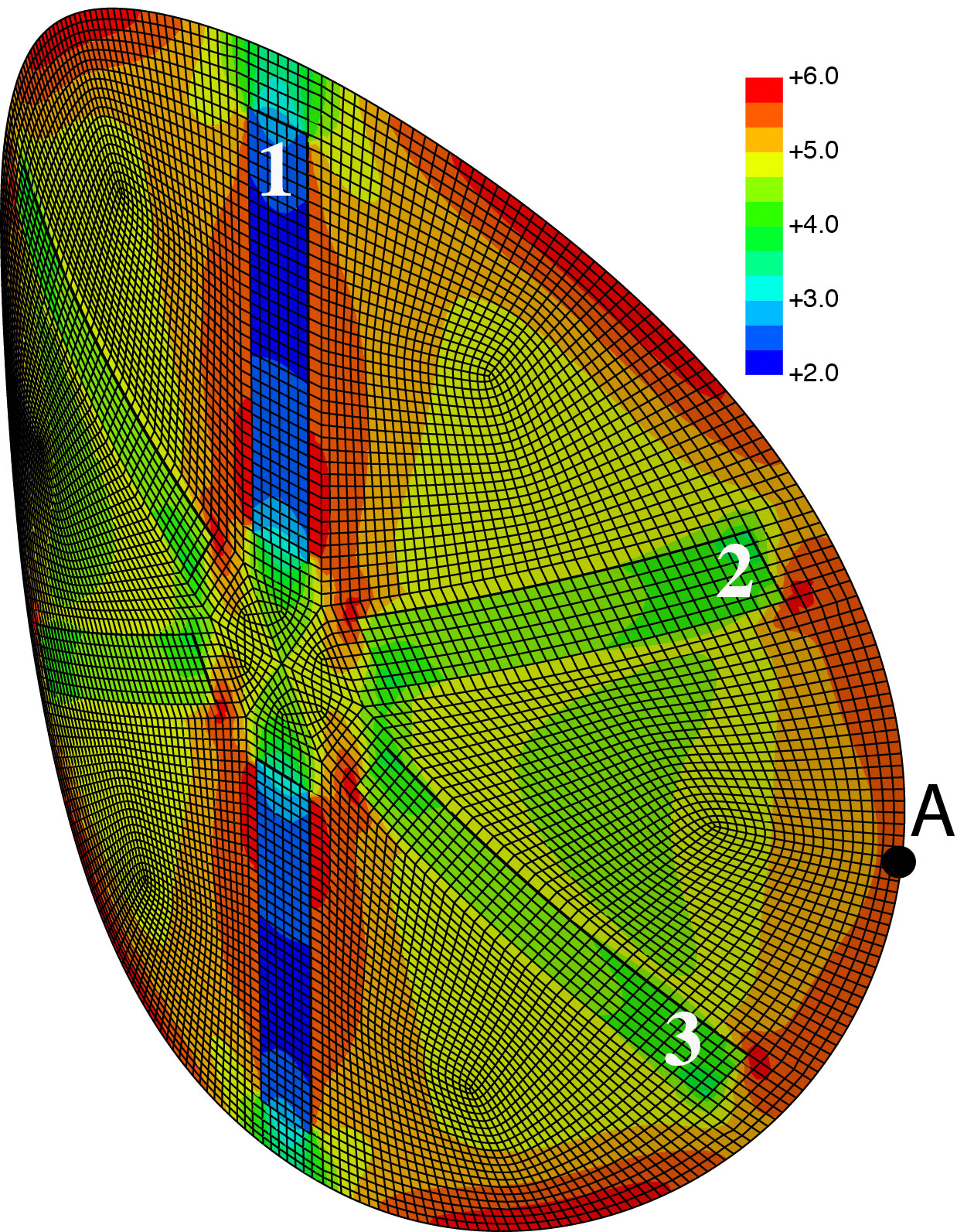}
\\
(c) & (d)
\end{tabular}
\caption{(a-b): Numerical diagrams to identify the non-dimensional shear stiffness $\alpha$ from the values of the critical mass $m_c$ for the stability loss and of the inelastic curvature $\bar k$ (see Appendix for details) without (a) and with MFC actuators (b). The contours are for constant values of $\alpha$; the  red dots correspond to the identified values, whilst the gray areas indicate approximative confidence intervals, see Table~\ref{tab:identification}. (c): Actuation test with $V_1=0$, $V_2=V_3=\bar V\in[0,1500] \mathrm{Volt}$  comparing the experimental measures of the transverse displacement of the point $A$ in (d) with the results of the uniform curvature model (UC) and a shell finite-element model implemented in \texttt{Abaqus} (FE). (d): Mesh and deformed configuration of the FE model in the same actuation test; the colours are the values of the curvature $k_{11}$.}
\label{fig:identification}
\end{figure}

\begin{table}[htp]
\caption{Measurements before and after bonding the  Micro-Fiber-Composite (MFC) actuators. }
\begin{center}
\begin{tabular}{lccccccc}
\toprule
 \multicolumn{1}{c}{\footnotesize }  &  \multicolumn{3}{c}{\footnotesize Measurements}  & \multicolumn{2}{c}{\footnotesize Max curvatures}&\multicolumn{2}{c}{\footnotesize Model parameters} \\
                    & $w^{(a)}$&$w^{(b)}$& $m_c$&$k^{(a)} $ & $k^{(b)}$&$\alpha$& $\bar{k}_\textsc{P}$\\\hline
{\footnotesize Without MFC}&$4.1\,\mathrm{cm}$&   $4.0 \,\mathrm{cm}$&   $32\pm5\,\mathrm{g}$          &               $5.7\,\mathrm{m}^{-1}$ &$5.5\,\mathrm{m}^{-1}$ & $1.08\pm0.02$&$4.2\pm 0.1\,\mathrm{m}^{-1}$\\
{\footnotesize With MFC  }   &$3.2\,\mathrm{cm}$ &   $3.1  \mathrm{cm}$&  $28\pm5\,\mathrm{g}$            &             $4.4\,\mathrm{m}^{-1}$&$4.2\,\mathrm{m}^{-1}$&$1.08\pm0.02$&$3.3\pm 0.1\,\mathrm{m}^{-1}$\\
\bottomrule
\end{tabular}
\end{center}
\label{tab:identification}
\end{table}%

\subsection{Plastically deformed disk with MFC actuators}
\label{sec:expMFC}
To actively control the  inelastic curvature we use commercially available  Macro-Fiber-Composite (MFC) actuators  from \cite{smartmaterial}, model \texttt{8514P1}, with dimensions $85\times14\times0.3 \mathrm{mm}$. They are composed of piezoelectric fibres embedded in an epoxy matrix and exploit the so-called $d_{33}$ piezoelectric coupling effect under operating voltages ranging from  $-500 V$ to $1500 V$. 
We bonded the MFC patches in the configuration shown in Figures~\ref{fig:configurationpzt}, while keeping the plastically deformed shell in a flat configuration. We use the specific glue  \texttt{Loctite M121HP} and a curing period of 48 hours at ambient temperature. 
The pairs of opposite patches along one direction are electrically connected in parallel and driven by the same electric voltage, \emph{i.e.}~each pair acts as a single actuator along its direction. We denote by $V_i$ the voltage of the pair aligned with the direction $\mathbf{e}( {\varphi}_i)=\{\cos {\varphi}_i, \sin{\varphi}_i\}$, with $i=1,2,3$.

The effects of the MFCs  are to (i)  increase the bending stiffness of the disk and, more importantly, (ii) introduce an additional inelastic curvature controlled by the applied voltages.
 
To theoretically estimate the parameters of the uniform curvature model \eqref{minrhouc} for the disk with the piezoelectric actuators, we combine Classical Laminate Plate Theory (CLPT, \cite{Red04}) with the  hypotheses of uniform curvature and inextensibility.
The MFCs can be considered as orthotropic piezoelectric layers, with  elastic properties 
$ E_1= 30.3 \,\mathrm{GPa}$, $E_2= 15.9 \,\mathrm{GPa}$, $\nu_{12} = 0.31$, $G_{12} = 5.5 \,\mathrm{GPa}$, where $1$ is the direction aligned with the longest side of the MFC, see \cite{smartmaterial}. With these values and with the properties \eqref{CPLTdisk} for the copper disk, we obtain the following updated parameters for the bending stiffness matrix and the equivalent inelastic curvature after bonding the MFC actuators\footnote{Explicit calculations of the parameters are reported in the directory \textsf{FromCLPTtoUCprecession} in the Supplementary Material.}
\begin{equation}
D = 0.0183\, \mathrm{N\,m^3}, \quad \beta=1,\quad  \nu = 0.329,\quad \alpha = 1.065, \quad \bar{k}_\textsc{P}=3.28\mathrm{m}^{-1}.
\label{CPLTMFC}
\end{equation}
Interestingly, the shell remains globally square-symmetric. Its anisotropy, measured by the ratio $\alpha$, is only slightly affected by the placement of the PZT patches: hence the energetic gaps between the equilibria is still small, see Figure~\ref{fig:coneisoaniso}a. The reduction of the equivalent inelastic curvature after bonding the MFCs is due to the increased stiffness of the shell and the fact that the MFCs are naturally flat. To validate these predictions, we repeat the same identification procedures of Section~\ref{sec:exp}\ref{sec:plastification} after bonding the MFCs. The results are shown in the second row of Table~\ref{tab:identification} and are in a good agreement with \eqref{CPLTMFC}.  The equilibrium curvatures are significantly reduced, but  the shell is still bistable, see Figure~\ref{fig:configurationpzt}, and largely verifies the inextensibility condition $k\sim 40 \,k^* \gg k^*$. 

The piezoelectric effect inside each MFC actuator can be modelled through inelastic strains proportional to the applied voltage in the form $\bar{\varepsilon}_{11}= \delta_{11} V$, $\bar{\varepsilon}_{22}= \delta_{12} V$, $\bar{\varepsilon}_{12}=0$, 
with $\delta_{11}=0.72\times 10^{-6}\mathrm{V}^{-1}$, $\delta_{12}=-0.38\times 10^{-6}\mathrm{V}^{-1}$, see \cite{smartmaterial}. Applying CLPT, we find that the equivalent inelastic curvature induced by the MFC actuators in Voigt notation is 
\begin{equation}
  \bar{\mathbf{k}}_{\textsc{V}}(t)=\sum_{i=1}^3\bar{\boldsymbol{\chi}} (\varphi_i)\,V_i(t),\quad
\bar{\boldsymbol{\chi}} (\varphi_i)=
 \left(
\begin{array}{c}
 \chi_1 \,\cos ^2\varphi_i+\chi_2 \,\sin ^2\varphi_i \\
\chi_1 \,\sin ^2\varphi_i +\chi_2 \,\cos ^2\varphi_i \\
(\chi_1-\chi_2)\, (\sin 2 \varphi_i) /\alpha 
\end{array}
 \right)
 \label{barkv}
\end{equation}
 where $\chi_1=-2.46 \,10^{-4}\,\mathrm{m^{-1} V^{-1}}$ and $\chi_2=+1.07 \,10^{-4}\,\mathrm{m^{-1} V^{-1}}$ are electromechanical coupling coefficients representing the curvatures per unit voltage in the directions parallel and perpendicular to the axis of the actuator, respectively.
We perform a  dedicated test to check the estimates of the piezoelectric couplings: with the shell initially positioned in the equilibrium configuration  of Figure~\ref{fig:configurationpzt}a with $\varphi=0$,  we connect in parallel the pairs $2$ and $3$, imposing a voltage $V_2=V_3=\bar{V}$, while keeping $V_1=0$. Figure~\ref{fig:identification}c reports the  measured displacement  of the peripheral point $A$ of the shell (see Figure~\ref{fig:identification}d)  when varying $\bar V$ in the range $[0,1500]\,\mathrm{Volt}$ , and compares it with the results of the uniform curvature (UC) model with the parameters \eqref{CPLTMFC}-\eqref{barkv}  and of a finite-element (FE) simulation.
The finite-element model of the disk with MFC actuators is implemented using the commercial code \texttt{Abaqus}\footnote{The \texttt{Abaqus} scripts arte reported in the directory \textsf{Abaqus} of the Supplementary Material.}. We use a fully nonlinear shell model (S4R elements) and account for the plastification and the piezoelectric effect through equivalent thermal inelastic strains, with the three-dimensional geometry and material parameters as inputs. Figure~\ref{fig:identification}d shows the mesh and the current configuration of the shell with the plastic curvature  and  $V_2=V_3=1500\,\mathrm{Volt}$. 
Overall, the agreement between the experimental results, the UC model, and the FE model is satisfactory, considering the complexity of the physical system and the simplicity of the UC model.  The differences between the FE model and the experimental results can be attributed to material nonlinearities of the MFC actuators at high voltages and to the effect of the bonding layer, which is neglected in the FE and UC models. The errors between the UC and the FE models can be explained by the fact that the UC model cannot account for  non-uniform curvatures.

\subsection{Piezoelectric controlled precession of the disk curvature axis }\label{sect:expactuation}
To obtain an inelastic curvature varying in time as in \eqref{actuationlaw} and Figure~\ref{fig:cases}, we impose the following voltages on the three MFC pairs  
\begin{equation}
V_1(t)=\dfrac{\bar{V}}{2} (1+\cos t),
\quad
V_2(t)=V_1(t-{2\pi}/{3}),\quad
V_3(t)=V_1(t+{2\pi}/{3}),
\label{cardioids}
\end{equation}
which are in the form of a rotatory travelling wave parametrised by the amplitude  $\bar{V}$ and the phase $t$.
The allowable voltages are in the interval $[-500,1500]\,\mathrm{Volt}$, but the behaviour of the MFC actuators is not symmetric for positive and negative voltages. Therefore, with \eqref{cardioids}, we have chosen to keep the voltages always positive. 
According to \eqref{barkv}, the corresponding piezoelectric induced inelastic curvature is 
\begin{equation}
 \bar{\mathbf{k}}_{\textsc{V}}(t)=
 \dfrac{3(\chi_1+\chi_2) \, \bar V }{4}
 \left(\begin{array}{c}
1\\1\\0 
\end{array} \right)+
\dfrac{3(\chi_2-\chi_1) \, \bar V }{8}
 \left(\begin{array}{c}
\cos t\\ -\cos t\\ \dfrac{2\sin{t}}{\alpha} 
\end{array}\right).
 \end{equation}
Whilst the first term is an isotropic component and  adds a negligible perturbation to initial plastic curvature, $\bar k_\textsc{P}\gg (\chi_1+\chi_2)\bar V$, the second term introduces the 
 desired precession of the principal axis of the inelastic curvature tensor as in \eqref{actuationlaw}. 
 With these relations, one can easily find that the thresholds $\bar\rho^*$ and $\bar\rho^{**}$ in \eqref{rhostar} on the  inelastic curvature translate to the following thresholds on the actuating amplitude   $\bar{V}$:
 \begin{equation}
 \bar V^{**}\simeq 2 \bar V^{*}\simeq
  \frac{4 (\alpha -1) (1+\nu)\,\bar k_\textsc{P}}{3 (\chi_2-\chi_1)}\simeq 1300\pm200 \,\mathrm {Volt}.
  \label{thresholdUC}
  \end{equation}
For the sake of simplicity, these expressions are reported in their linearised version for $\alpha$  close to $1$; the numerical value corresponds to the parameters reported in Section~\ref{sec:exp}\ref{sec:plastification}-\ref{sec:expMFC}, and the  confidence interval roughly accounts for the uncertainty in the estimation of these parameters.

 \begin{figure}[h]
\begin{center}
\subfloat[]{
\includegraphics[width=.8\textwidth]{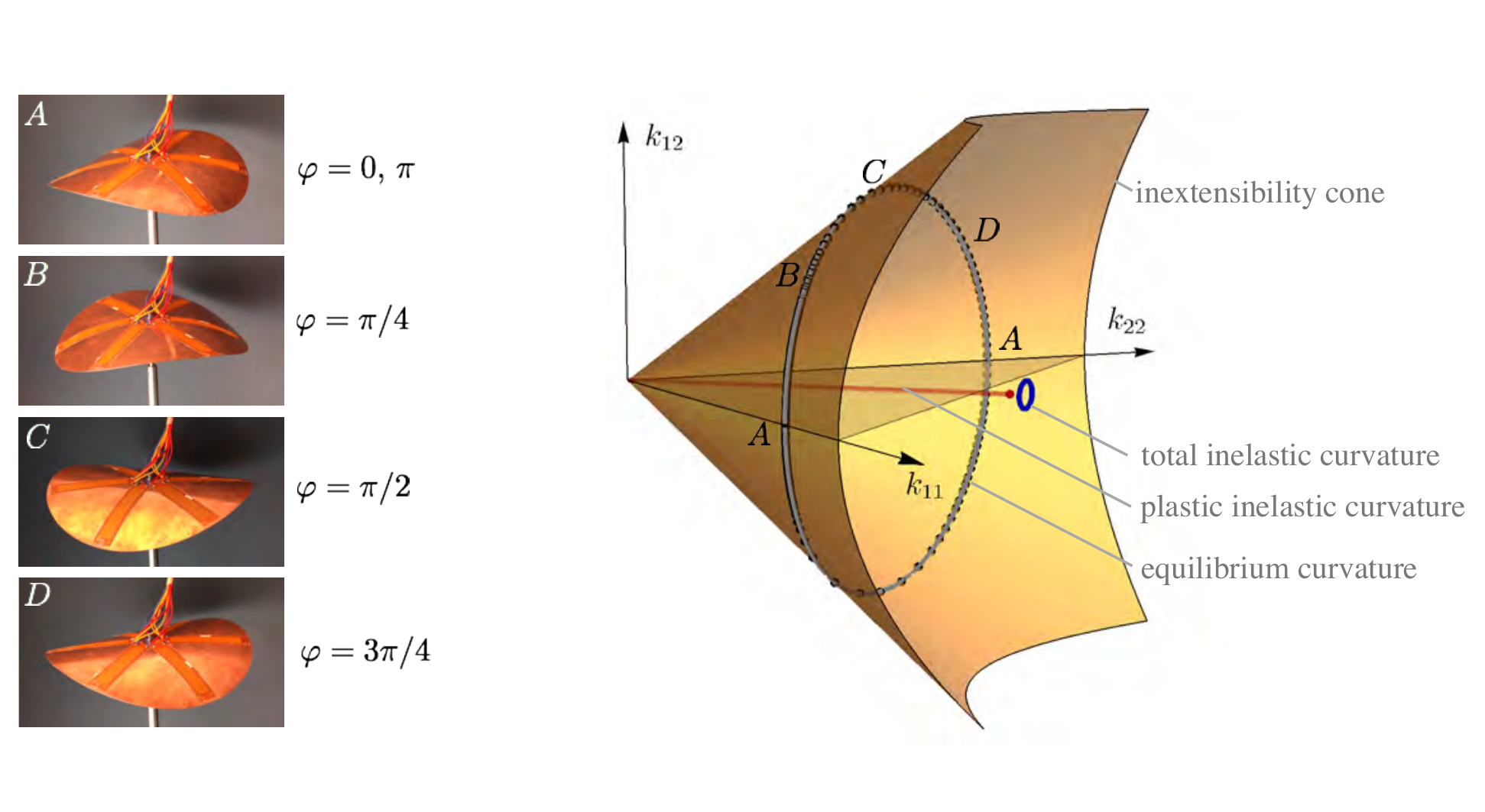}}
\\
\subfloat[]{
\includegraphics[height=0.4\textwidth]
{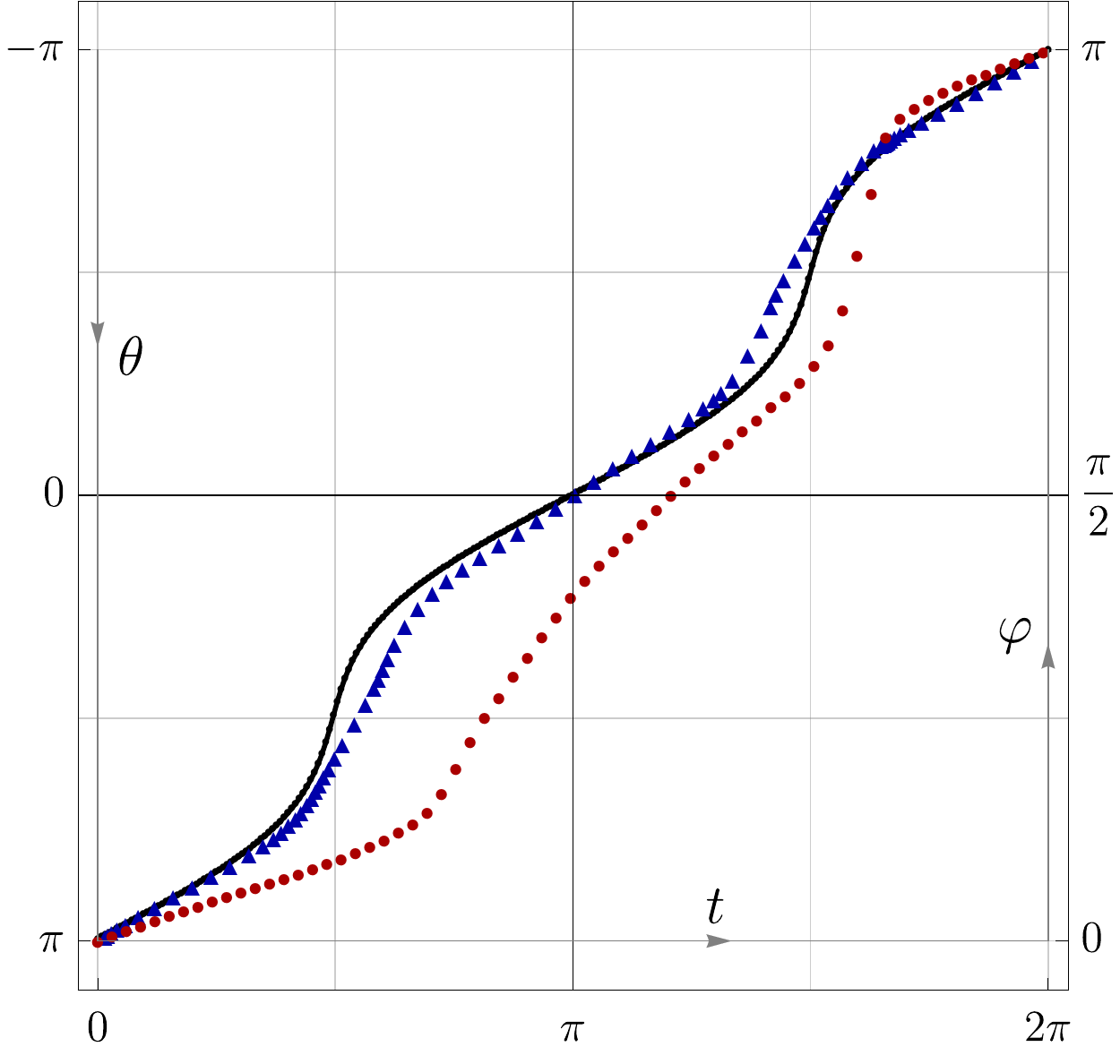}
}
\subfloat[]{
\includegraphics[height=0.4\textwidth]{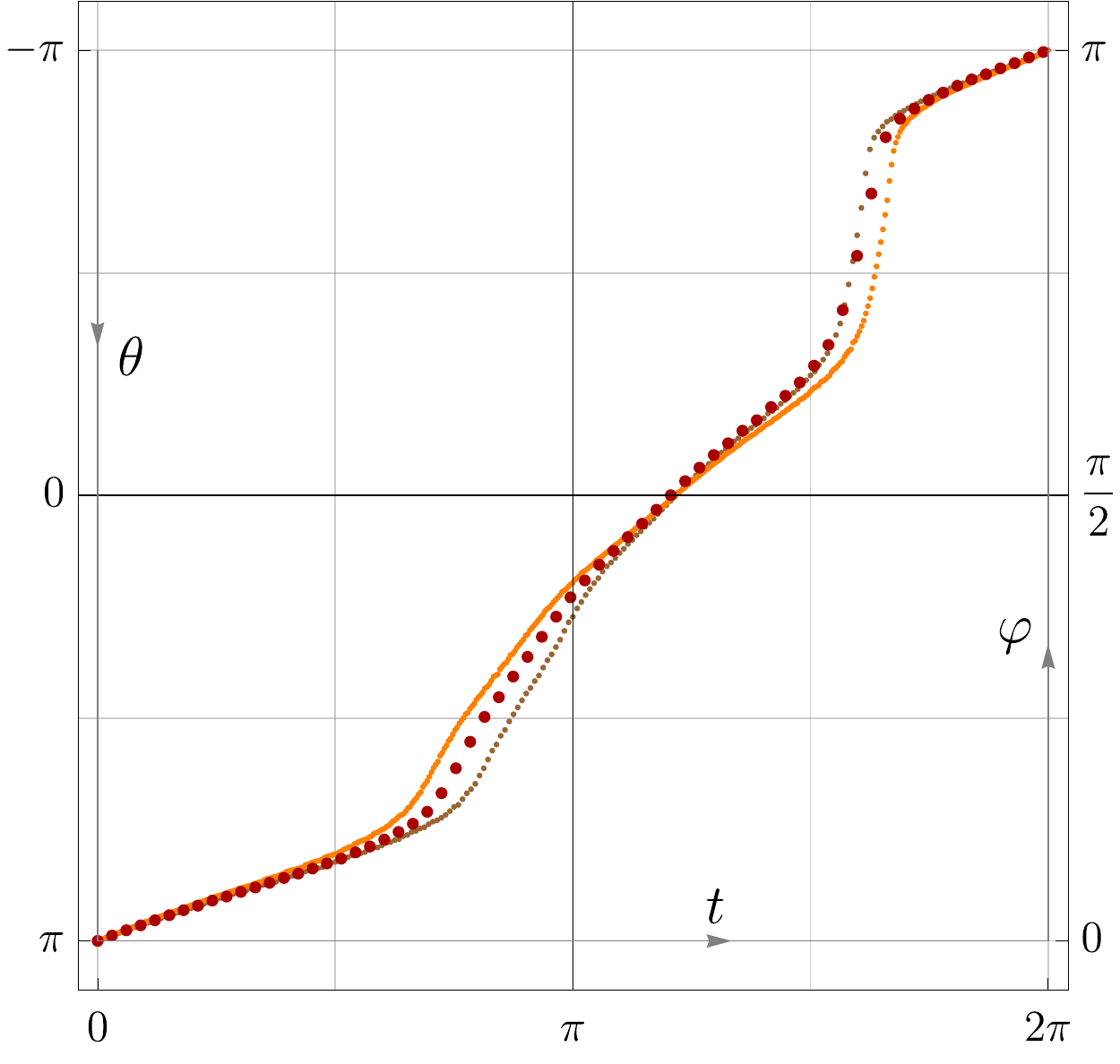}
}\end{center}
\caption{Precession of the curvature axis of the disk controlled through the multiparameter piezoelectric actuation. (a):  Loci of inelastic curvature (small blue ellipse) and resulting  curvature at the equilibrium (continuous line: UC; points: experiments) when applying the actuation law \eqref{cardioids}. The  cone represents the inextensibility constraint $\det \mathbf k=0$. The insets, A to D, display the shell configurations observed experimentally. (b) and (c): Orientation of the axis of maximal curvature  ($\theta$ or $\varphi$) when varying the  inelastic curvature \eqref{actuationlaw} for $t$ in $[0,2\pi]$. (b): Comparisons between experimental result (red dots,  $\pi/150\, \mathrm{rad/s}$) and the predictions of the FE  (blue triangles) and UC (black line) models. (c): Experimental results for the actuation rates $\pi/30$, $\pi/90$ and $\pi/150\,\mathrm{rad/s}$.}
\label{fig:sequencecone}
\end{figure}
The analysis of Section~\ref{sect:act} based on the uniform curvature model predicts  a complete stable quasi-static precession of the axis of maximal curvature of the shell when applying the actuation law \eqref{cardioids} with $\bar V >\bar V^{**}$, whilst for $\bar V^{*}< \bar V <\bar V^{**}$ the precession will include a snap-through phenomenon. For $\bar V < \bar V^{*}$ the shell will remain in a neighbourhood of the initial configuration. 

To experimentally validate the effectiveness of the proposed actuation law, we apply the voltages \eqref{cardioids} on the three MFC pairs by using the three-channel high-voltage amplifier HVA1500-3  by {Smart Material GmbH} \cite{smartmaterial}. The input signals are digitally generated in \texttt{Matlab} \cite{MATLAB:2016b} and transferred to the amplifier through a multi-channel D/A converter. The shell is initially clamped horizontally at its center onto a vertical rod using a strong locating magnet.
We  record the time evolution of the deformed shape through a standard camera giving a  top view of the shell. From the frames of the video we reconstruct, using \texttt{ImageJ} software \cite{ImageJ}, the orientation of the principal axis of curvature at each instant $t$.

Figure~\ref{fig:sequencecone} illustrates the main experimental results, anticipated in  Figure~\ref{fig:sequence}: when varying the phase $t$ of the actuation law~\eqref{cardioids}, the axis of maximal curvature  of the disk experiences a complete, quasi-static, precession (see also the Videos in Supplementary Materials). The rotation of the disk is only apparent, as evident from the invariant orientation of the MFC patches: the angular velocity of the material points is null. The results shown in Figure~\ref{fig:sequencecone} are for $\bar V=1500\mathrm{Volt}$. 
The minimal threshold to obtain the quasi-static precession is  $\bar V\simeq1200-1400\,\mathrm{Volt}$, a result in striking quantitative agreement with the theoretical prediction \eqref{thresholdUC} of the uniform curvature model. 
In Figure~\ref{fig:sequencecone}a, the small blue ellipse shows the variation of the inelastic curvature in the $(k_x, k_y, k_{xy})$ space when applying the actuation law \eqref{cardioids}. This inelastic curvature is composed of a large constant plastic component, the red line in Figure~\ref{fig:sequencecone}a, and a small additional piezoelectric contribution controlled by the applying voltage, which is varying in time. The large ellipse reports the associated average curvature tensors at the equilibrium, obtained either experimentally (black dots) or with the UC model (continuous line). The curvatures at the equilibrium respect to a large degree the inextensibility constraint $\det \mathbf k=0$, represented by the yellow cone.

Figure~\ref{fig:sequencecone}b-c gives the correspondence between the orientation $t$ of the inelastic curvature and the equilibrium curvature, $\varphi$ (or $\theta$).  Figure~\ref{fig:sequencecone}b compares the experimental results to the prediction of the FE and UC models. The FE and UC models are in very close agreement.
Dedicated simulations, not reported here, show that the discrepancies with the experimental results and, in particular, the fact that in the experiment $\theta>\pi$ for $t=\pi$ could be explained by the effect of nonvanishing $D_{13}$, $D_{23}$ terms in the bending stiffness matrix (non-orthotropy). Figure~\ref{fig:sequencecone}c reports the experimental results  obtained when  varying the angular velocity of the rotation of the inelastic curvature axis from $\pi/30$ to $\pi/150$ $\mathrm{rad/s}$, showing that, in this regime, the phenomenon is almost rate-independent.

\section{Conclusions}\label{sect:concl}
We proved a novel concept to effectively control large structural shape-changes through a weak multiparameter piezoelectric actuation. The key result is outlined in Figure~\ref{fig:sequencecone}$a$: by plastically deforming an initially flat disk, we conceived  a cylindrical shell where  the embedded piezoelectric actuation generates small perturbations of the initial inelastic curvature but induces a complete precession of the shell curvature axis. The device behaves as a \emph{gearless motor}, where the multiparameter piezoelectric actuation generates a large-amplitude travelling wave for the transverse displacement. 

With standard bending actuation the maximum structural displacements induced by the piezoelectric actuation are of the order of $2$~mm (see Figure~\ref{fig:identification}$c$), for a disk of thickness $0.3$~mm and radius $12$~cm. With the proposed actuation technique, we are able to obtain displacements of   $3.2$~cm; such a displacement can actually be increased by increasing the inelastic curvature $\bar k_\textsc{P}$ or the disk radius, but the forces exerted in quasi-static conditions remain small, see \eqref{stabmarginforf} in the Appendix for an order of magnitude. More in general, while with standard actuation techniques the achievable curvature/voltage  ratio is of the order of $\chi_i$, leveraging the vanishing stiffness mode the proposed strategy induces a variation of the curvature per unit voltage of the order of 
\begin{equation}
{\bar k_\textsc{P}}/{\bar V^{**}}\sim{ \chi_i}/{\vert\alpha-1\vert},
\end{equation}
as shown by \eqref{thresholdUC}.
The amplifying factor is theoretically infinite for an ideal perfectly isotropic, neutrally stable, shell.

We studied   the specific case of an almost neutrally stable disk obtained after a suitable plastification process. The use of a fully nonlinear inextensible shell model  based on the uniform curvature hypothesis allowed us to get a full understanding of the possible equilibrium shapes and to design the appropriate multi-parametric actuation law. We showed that the key design parameter is the deviation of the shear stiffness, $\alpha $ in \eqref{Dmatrix}, from the one of an isotropic shell. We were able to control this parameter in experiments and theoretical models, showing how it determines the actuation threshold required to obtain the precession of the curvature axis \eqref{thresholdUC}. 

We believe that this work can provide new ideas to address the important technological problem of controlling the shape of structures through embedded active materials, a fundamental issue of flexible robotics. The concept illustrated here can be extended to more complex structures, such as helical shells \cite{GuoMehGro14} and multistable strips \cite{GioMah12}.  We fabricated an experimental prototype of sub-metre scale. However, the key concepts are scale invariant and the same ideas can be applied to {micro-electro-mechanical} systems (MEMS). In MEMS the elastic mismatch, arising during the deposition process of thin films, can replace the role of  the initial plastic curvature \cite{FreSur04}. 
Further interesting perspectives are the applications to flexible robot locomotion  \cite{desimone2015,ArrHelMil12} and the nonlinear energy harvesting exploiting the direct piezoelectric effect \cite{arrieta2010}.

\section*{Appendix}
\label{sec:app}
In this Appendix we derive the stability threshold of the equilibrium configurations $\theta=0,\pi$ when applying transverse forces $f$ as sketched in the insets of Figures~\ref{fig:identification}a-b. This calculation is used to identified the material parameter $\alpha$ in Table~\ref{tab:identification}.

The total energy $\mathcal{E}_t$ is obtained subtracting the work of external forces $\mathcal{L}_f$ to the elastic energy $\mathcal{E}_b=2 D U(\kappa_m,\theta)$:
\begin{equation}
\mathcal{E}_t(\kappa_m,\theta)= 2 D \, U(\kappa_m,\theta) -\mathcal{L}_f,\qquad \mathcal{L}_f:=- 2 f L^2 \kappa_m \dfrac{1+\cos\theta}{2\sqrt{1+\nu}},
\label{Utot}
\end{equation}
where $U(\kappa_m,\theta)$ is given in \eqref{mincone2}. For $\alpha\ge 1$, there exist equilibria $\theta=0, \pi$ with 
\begin{equation}
\kappa_m=
\dfrac{\left(\sqrt{\beta }+\nu \right)}{2 D
   \sqrt{\nu +1} } \cdot
   \dfrac{f L^2 (1+\cos\theta)-2 D \bar{\kappa}_m \sqrt{\nu +1}}{\left(\sqrt{\beta }-\nu \right)
   \left(\alpha  \sin ^2\theta +\cos ^2\theta
   \right)+\sqrt{\beta }+\nu}.
\label{kmsolwithf}
\end{equation}
Inspecting the Hessian matrix of second derivatives of the energy, we find that the equilibrium $\theta=0$ is stable for 
\begin{equation}
\left(f L^2-D \sqrt{\nu +1} \kappa _m\right) \left(\frac{(\alpha
   -1) \left(\sqrt{\beta }-\nu \right) \left(2 f L^2-2 D
   \sqrt{\nu +1} \kappa _m\right)}{2 \sqrt{\beta }}+f L^2\right)\ge 0.
\label{Utotstability}
\end{equation}
The critical values of the forces, leading to a vanishing stability margin, are, then, found to be 
\begin{equation}
f_{c1}=\frac{D \bar\kappa _m \,\sqrt{1+\nu} }{L^2}\cdot  \dfrac{(\alpha -1)  \left(\sqrt{\beta }-\nu
   \right) }{\alpha  \left(\sqrt{\beta }-\nu
   \right)+\nu },\quad
   f_{c2}= \frac{D \bar\kappa _m \,\sqrt{1+\nu} }{L^2},
\label{stabmarginforf}
\end{equation}
where for $\alpha\ge 1$, $f_{c1}\le f_{c2}$. Therefore, one can estimate $\alpha$ measuring the lowest critical value of the force $f$ leading to instability. We obtain
\begin{equation}
\alpha\simeq 1+\dfrac{ f_{c1} L^2 \sqrt{\beta } (1-\nu)}{\bar{k}_m Y t^3
   \left(\sqrt{\beta }-\nu \right)}.
\label{alphaid}
\end{equation}

\dataccess{The datasets supporting this article have been uploaded as supplementary material.}
\aucontribute{W.Hamouche performed the experiments. All the authors contributed in equal measure to the theoretical and numerical work, and in the writing of the manuscript. All the authors gave final approval for publication.}
\competing{We have no competing interests.}
\funding{The authors acknowledge the financial
support of Project ANR-13-JS09-0009 (Agence Nationale de la Recherche).}
\ack{The authors acknowledge Benoit Roman (ESPCI), Francois Ollivier, and Jo\"el Pouget (UPMC) for the support in the experimental work.}
\bibliographystyle{unsrt}

\end{document}